\documentclass[useAMS,usenatbib,babel]{mn2e}
\usepackage{mathpazo}

\usepackage[english,english]{babel}
\usepackage{amsmath}
\usepackage{amssymb,amsfonts,textcomp}
\usepackage{array}
\usepackage{supertabular}
\usepackage{hhline}
\usepackage{hyperref}
\usepackage[usenames]{color}
\hypersetup{dvips, colorlinks=true, linkcolor=blue, citecolor=blue, filecolor=blue, urlcolor=blue}
\usepackage[dvips]{graphicx}

\def\gtrsim{\lower.5ex\hbox{$\; \buildrel > \over \sim \;$}}
\usepackage{graphicx}

\newcommand{\msun}{\mbox{$M_\odot$}}
\newcommand{\mvir}{\mbox{$M_{\rm vir}$}}
\newcommand{\rvir}{\mbox{$R_{\rm vir}$}}
\newcommand{\mn}{\mbox{{\sc \small Horizon}-MareNostrum\,\,}}
\newcommand{\nut}{\mbox{{\sc \small Nut}}}

\begin{document}

\author[C. Pichon, D. Pogosyan, T. Kimm,  A. Slyz, J. Devriendt and Y. Dubois ]{C. Pichon$^{1,2,3}$, D. Pogosyan$^{4}$, T. Kimm$^2$, A. Slyz$^2$, J. Devriendt$^{2,3}$ and
Y. Dubois$^{2}$  \\
$^{1}$ Institut d'Astrophysique de Paris, 98 bis boulevard Arago, 75014 Paris, France\\
$^{2}$ Astrophysics, University of Oxford, Keble Road, Oxford OX1 3RH, UK.\\
$^3$ Observatoire de Lyon (UMR 5574), 9 avenue Charles Andr\'e, F-69561 Saint Genis Laval, France. \\
$^{4}$Department of Physics, University of Alberta, 11322-89 Avenue, Edmonton, Alberta, T6G 2G7, Canada.
 }

\title[Rigging dark halos]{
{Rigging dark halos:}
{why is hierarchical galaxy formation consistent  with the inside-out build-up of thin discs?}
}

\maketitle

\begin{abstract}
{State-of-the-art hydrodynamical simulations show that gas inflow through the virial sphere of dark matter halos is focused (i.e. has a preferred
inflow direction), consistent (i.e. its orientation is steady in time) and amplified
(i.e. the amplitude of its advected specific angular momentum increases with time).
We explain this to be a consequence of the dynamics of the cosmic web within the
neighbourhood of the halo, which produces steady, angular momentum rich, filamentary inflow of cold gas. On large scales, the dynamics
within neighbouring patches drives matter out of the surrounding voids, into walls and
filaments before it finally gets accreted onto virialised dark matter halos. As these walls/filaments constitute the boundaries of asymmetric voids, they
acquire a net transverse motion, which explains the angular momentum rich nature of the later infall which comes from further away.
We conjecture that this large-scale driven consistency explains why cold flows are so efficient at building up high redshift thin discs from the inside out.}
\end{abstract}

\section{Introduction}

{
{One of the persistent puzzles of the standard paradigm of galaxy formation is the following:
why do we observe
}{\textit{thin}}{ galactic discs when hierarchical clustering naively suggests
these galaxies should undergo repetitive random interactions with satellites and incoherent gas infall from their environment?
Indeed whilst one can argue that the probability of a head on
collision with satellites should be small,
incoherent but continuous gas infall poses a much greater threat to
the ability of a bottom-up scenario of galaxy formation to form ubiquitous thin galactic discs.}}
{
Historically, astronomers \citep{rees77,silk77} have invoked the primordial
monolithic collapse of a spheroidal body of gas which is shock-heated to its
virial temperature. In this scenario, subsequent in-falling gas \citep[now described as secondary infall]{fillmore84,bert85}
shock-heats as it hits the virial radius, while the inner, denser
region of the hot gaseous sphere secularly rains onto the central disc
as it radiatively cools.
This process has been coined the ``hot mode" of gas accretion and implies a clear correlation between
the spin of the hot galactic corona and that of the disc which is effectively assumed to be shielded from its cosmic
environment \citep{fall80,mo98,bullock01}.}

{
Over the last few years it has been (re)realized \citep{birnboim03,keres05,ocvirk08,brooks09,dekel09} that
most of the gas feeding galaxies arrives cold ($\approx$ 10$^4$K) from the large scale structures along filaments, and does
not create such a halo filling corona of hot gas, as the radiative shock it
must necessarily generate to do so is unstable to cooling processes. These investigations correspond
to an update in the current $\Lambda$-CDM cosmological framework of the basic prediction of
\citet{binney77} who argued that for all but the most massive
galaxies, the accreted gas, provided it was dense enough, would never shock-heat to temperatures where
Bremsstrahlung dominates cooling as it would first cool by atomic transitions.
In the author's own words, the accretion shock would
be `isothermal' rather than `adiabatic' and consequently, only a
negligible fraction of the gas would ever reach temperatures $T \sim T_{\rm vir}$.
In parallel to the work of \citet{aubert04} on anisotropic }{\textit{dark
matter}}{ infall onto galactic halos, \citet{katz03} and \citet{keres05}
numerically confirmed that a large fraction of
the gas is indeed accreted through filamentary streams, where it remains cold before it reaches the galaxy
(see also  \citealt{kay00,fardal01} for early indication that this gas was not
shock-heated). This ``cold-mode'' accretion dominates the global growth of all
galaxies at high redshifts ($z \geq 3$) and the growth of low mass (M$_{\rm
halo} \le 5 \times 10^{11}$ M$_\odot$) objects at late times \citep{dekel06}. }

{
The addition of  accretion through cold streams to the standard galaxy formation framework has
received much attention  \citep[e.g.,][]{keres09,brooks09}
because of its potential implications for the star formation history of
galaxies (although see \citealt{bensonbower11} for a defense of the opposite point of view). In
the traditional `hot mode' picture, star formation is delayed as accreted gas is shock-heated  and requires time to cool
onto the central object. In contrast, if this material comes in cold,
star formation can be fueled on a halo free-fall time. Cold-mode
accretion should also have an important impact on the properties
(scale length, scale height, rotational velocity) of galactic discs, if as
conjectured by \citet{keres05},  cold streams merge onto disks ``like streams of cars
entering an expressway'', converting a significant fraction of their infall velocity to
rotational velocity. \citet{dekel09} argued along the same lines
in their analysis of the \mn simulation: the stream carrying the largest coherent flux with an impact parameter of
a few kiloparsecs may determine the disc's spin and orientation. \citet{powell10} spectacularly confirmed these conjectures
by showing that indeed, the filaments connect rather smoothly to the disc: they appear to join from different
directions, coiling around one another and forming a thin extended disc structure, their high velocities driving its rotation. }

{
The way angular momentum is advected through the virial sphere as a function of time is expected to play a key role in
re-arranging the gas and dark matter within dark matter halos. The pioneer works of  \citet{peebles69,doroshkevich70,white84} addressed the
issue of the original spin up of collapsed halos, explaining its linear growth up to the time the initial overdensity
decouples from the expansion of the Universe through the re-alignment of the primordial perturbation's inertial tensor with the shear tensor.
However, little theoretical work has been devoted
to analysing the outskirts of the Lagrangian patches associated with virialised
dark matter halos, which account for the later infall
of gas and dark matter onto the already formed halos.
In this paper, we quantify how significant this issue is and present a consistent picture of the time evolution
of angular momentum accretion at the virial sphere based on our current theoretical understanding of the large scale structure dynamics.
More specifically, the paper presents a possible answer to the conundrum of why cold gas flows in $\Lambda$-CDM universes are consistent with
thin disk formation. Indeed, as far as galactic disc formation is concerned, the heart of the matter lies in understanding
how and when gas is accreted through the virial sphere onto the disc. In other words,
what are the geometry and temporal evolution of the gas accretion?
}

{
In the 'standard' paradigm of disc formation, this question was split in two. The dark matter and gas present in the
virialised halo both acquired angular momentum through tidal torques in the pre-virialisation stage, i.e. until turn-around \citep[e.g][]{white84}.
The gas was later shock-heated as it collapsed,  and secularly cooled and condensed into a disk \citep{fall80} having lost most of the connection
with its anisotropic cosmic past. In the modern cold mode accretion picture which now seems to dominate all but the most
massive halos, these questions need to be re-addressed. This paper presents a new scenario in which the coherency in the disk build-up stems
from the orderly motion of the filamentary inflow of  cold gas coming from the outskirts of the collapsing galactic patch.
{
The outline is as follows: in section \ref{sec:hydro}, using hydrodynamical simulations, we report evidence
that filamentary flows advect an ever increasing amount of angular momentum through the halo virial sphere at redshift higher than 1.5. We also demonstrate
that the orientation of these flows is consistent, i.e. maintained over long periods of time. Section \ref{sec:story} presents results
obtained through simplified pure dark matter simulations of the collapse of a
Lagrangian patch associated with a virialised halo as
these have the merit of better illustrating the dynamics of matter flows in the outskirts of the halo.
Section \ref{sec:conclusion} is devoted to the presentation of the conjectured impact of this scenario
on disk growth at various redshifts, conclusions and prospects.
}

\section{Hydrodynamical evidence }
\label{sec:hydro}

{Let us start by briefly reporting the relevant hydrodynamical results we have obtained. We statistically analysed the
advected specific angular momentum of both gas and dark matter at the virial radius of dark haloes
in the \mn cosmological simulation at redshift 6.1, 5.0, 3.8, 2.5
and 1.5 (see Figure~\ref{fig:visual}, Details  can be found in \citealt{kimm11}).
\begin{figure}
   \includegraphics[width=8.5cm]{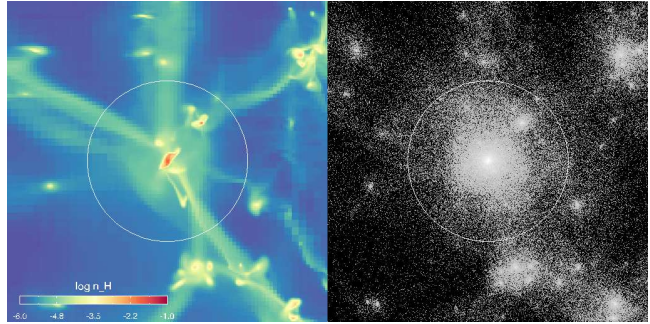}
    \caption{ 
A typical galaxy residing in a high mass halo ($M\sim 2 \times 10^{12}M_\odot$ at  $z=3.8$). 
The radius of the  circle in the both panels corresponds to $R_{\rm vir}=79$ kpc.
Gas ({\sl left panel}), and dark matter ({\sl right panel}) projected densities are plotted. 
Gas filaments are significantly thinner than their dark matter counterpart.
{Note the extent and the coherence of the large scale  gaseous filaments surrounding that galaxy.}
}
\label{fig:visual}
\end{figure}

\begin{figure*}
    \includegraphics[width=17cm]{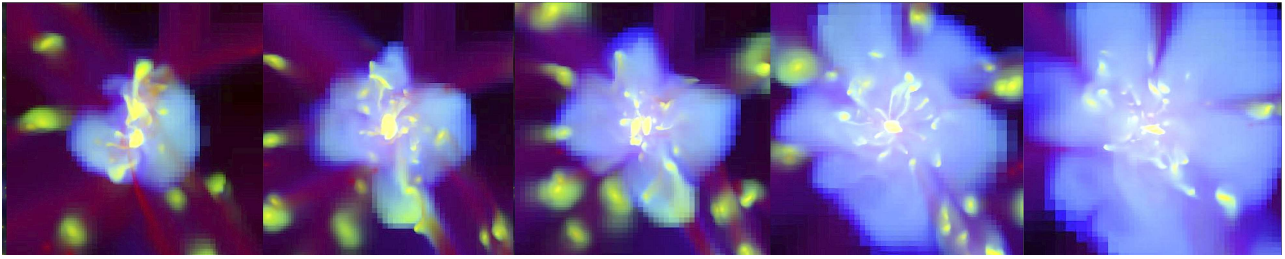}
    \caption{
{\sl These panels } display a time  sequence representing the evolution of the halo between $z=4$ and $z=2.5$
colour coded with density ({\sl red}) metallicity ({\sl green}), and temperature ({\sl blue}) within a fixed reference frame centred on the dark matter
halo at all times. The consistency of the direction of the infalling gas ({\sl in red}) is obvious.
}
\label{fig:visual2}
\end{figure*}

\begin{figure}
 \includegraphics[width=8.5cm]{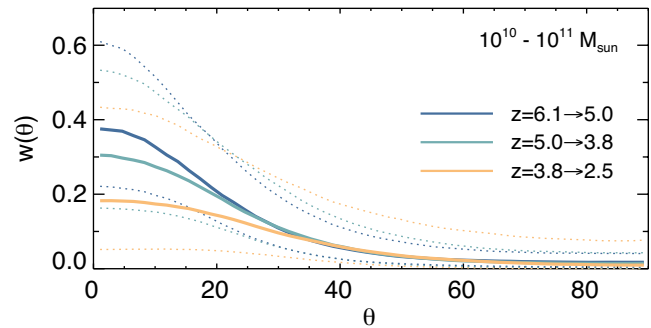}
\caption{ The covariances ({\sl thick line}) between different redshifts (as labeled) of
the thresholded density maps on the virial sphere, \rvir, together with
the corresponding dispersion (inter-quartile, {\sl dotted lines}). 
The lower bound of the thresholded density is chosen such that filamentary structures stand out,
while the upper bound is adopted to minimise the signal from the satellites (see the text, Section~\ref{sec:hydro}).
The orientation of filaments is temporally coherent, as is qualitatively illustrated in Figure~\ref{fig:visual2}.
}
\label{fig:correl-fil}
\end{figure}

\begin{figure*}
 \includegraphics[width=8.5cm]{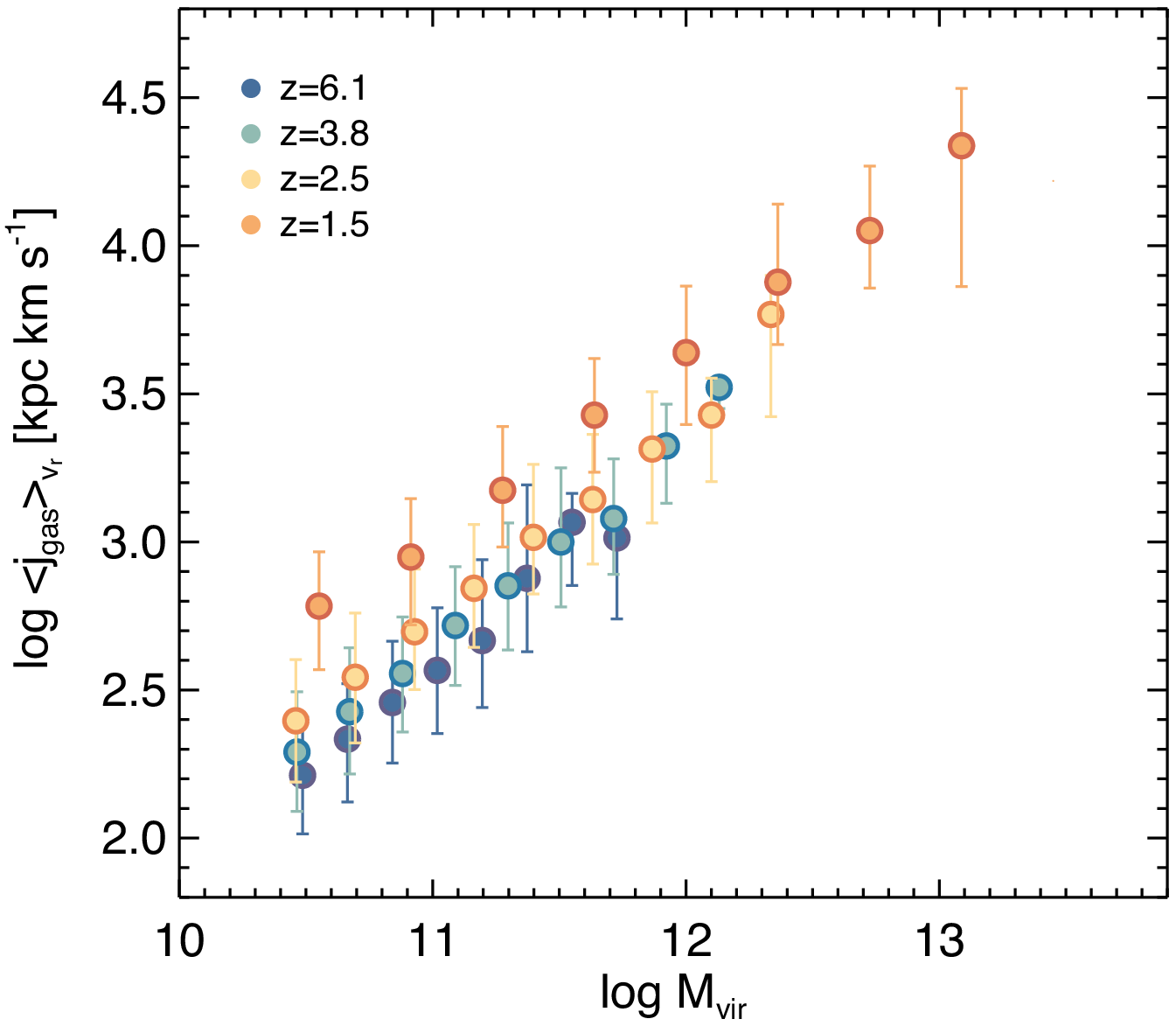}
\includegraphics[width=8.4cm]{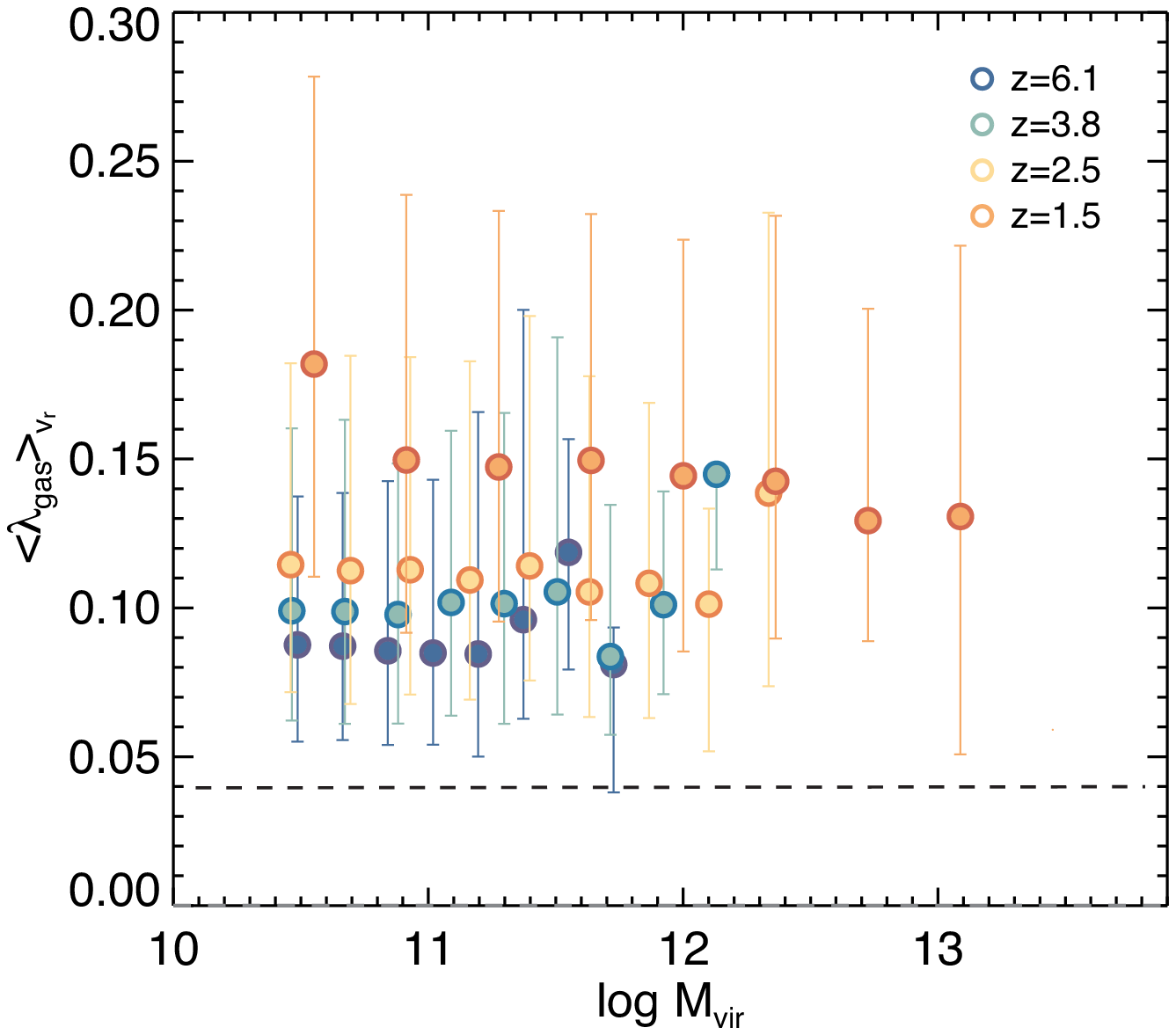}
\caption{{\sl Left}: advected specific momentum of gas as a function of cosmic time and
mass in the \mn simulation.
Note that more and more gas angular momentum is being advected as a function of time and mass.
{\sl Right}: \ advected spin parameter of gas as a function of cosmic time and mass;  a redshift trend persists, while the amplitude of
the advected spin parameter is larger than the averaged value of 0.04 (indicated by the horizontal dashed line) measured for dark matter halos
independently of time and mass \protect\citep{barnes87,kimm11}.}
\label{fig:jvr_M}
\end{figure*}

\begin{figure}
  \includegraphics[width=3.198in]{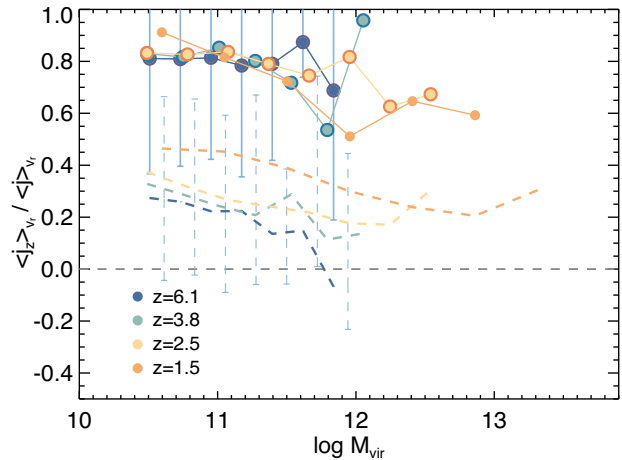}
\caption
{The ratio of the component of the advected gas's spin along the axis defined by the spin of dark
matter,  to the amplitude of the advected gas's
spin, as a function of mass and redshift, for the dense ({\sl
solid})/diffuse ({\sl dashed}) component
in the \mn simulation.
This measurement corresponds to the
median of 3119, 6655, 11357, 15419, and 15999 halos, resp. for $z=6.1$, 5.0. 3.8, 2.5, and 1.5. 
 The cold flow is more aligned than
the diffuse flow. This supports the view that the momentum rich gas flows along
the filaments and \ preserves most of its orientation. See
Figure~\ref{fig:example-jz} for a map illustrating the concentration of momentum in the filaments.
}
\label{fig:jz_M}
\end{figure}

\begin{figure}
 \includegraphics[width=8.75cm]{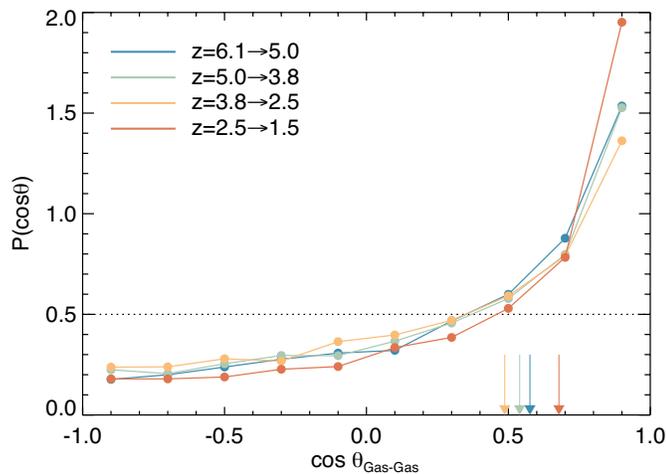}
\caption{
PDF of the cosine of relative angle between the spin of the advected gas,  $\langle { \mathbf j}\rangle_{v_r}$,
 for different pairs of  redshifts as labeled. The mean of this PDF is also represented as a vertical
arrow for the various pairs of redshifts considered.  The orientation of the spin of filaments is temporally
coherent. }
\label{fig:cosinePDF}
\end{figure}

\begin{figure}
 \includegraphics[width=8cm]{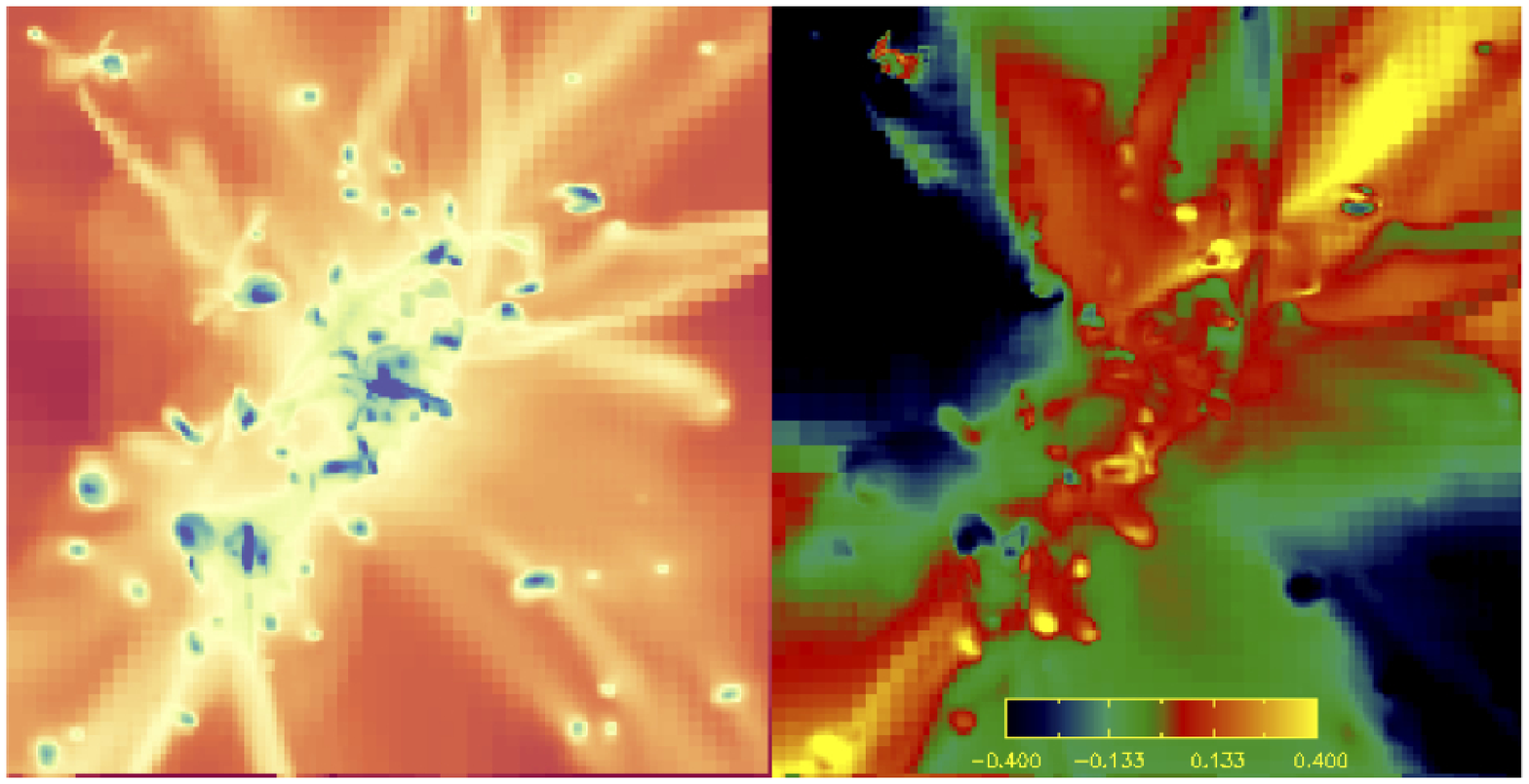}\\
\includegraphics[width=8cm]{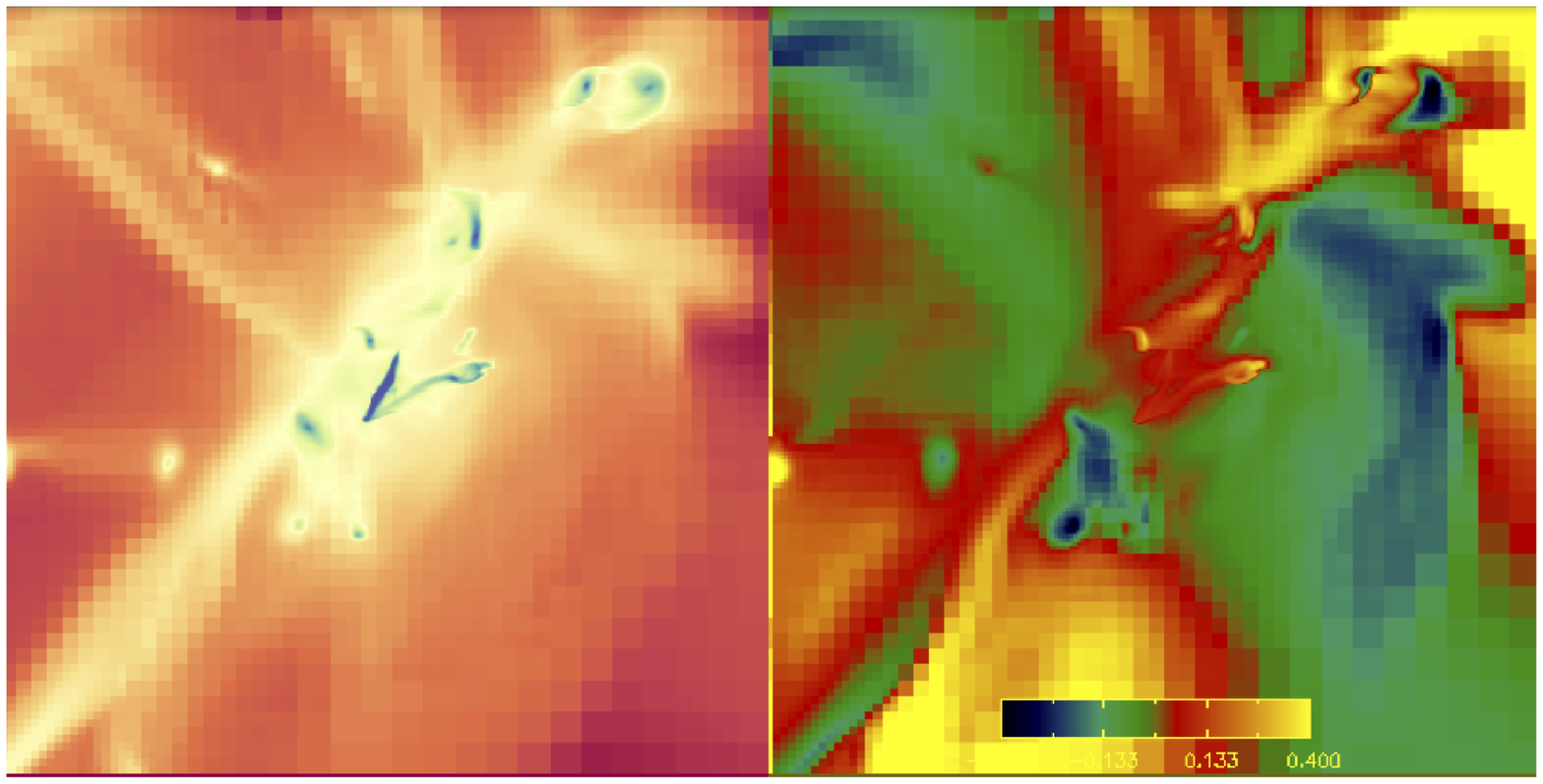}
\caption{ The distribution of the $z$ component (i.e. along the spin axis of their dark matter halo) of the spin
parameter ($\lambda_z$, {\sl right panels}) at z=3.8 in the \mn simulation for two different mass haloes: $5 \times 10^{12}$ ({\sl top panels}) and $3\times 10^{12} M_{\odot}$({\sl bottom panels}).
Projections of the hydrogen density are shown in the left panels.
The field of view is 4 virial radii on a side. It is apparent that
some filaments  display large values of   $\lambda_z$ (colour coded in yellow).}
\label{fig:example-jz}
\end{figure}

The \mn simulation \citep{ocvirk08,devriendt10} was carried out using the Eulerian hydrodynamic code,
{\sc \small ramses} \citep{teyssier02},
which uses an Adaptive Mesh Refinement (AMR) technique.
{It followed the evolution of a cubic cosmological volume of 50h$^{-1}$ Mpc on a side (comoving), containing
1024$^3$ dark matter particles and an Eulerian root grid
of 1024$^3$ gas cells. A $\Lambda$CDM
concordance universe ($\Omega_{m} = 0.3, \Omega_{\Lambda} =
0.7, \Omega_b=0.045, h=H_{0}/[100 \ {\rm km} s^{-1}
{\rm Mpc}\textsuperscript{-1}] =0.7, \sigma_{8} = 0.9, n = 1$)
corresponding to the WMAP 1 best fit cosmology  was adopted,
resulting in a dark matter particle mass $m_{p} = 1.41\times
10^{7} M_{{\sun}}$. A quasi Lagrangian refinement  policy was enforced to keep the
spatial resolution fixed at about 1 h$^{-1}$kpc in
physical coordinates. } A uniform UV background instantaneously turned on at $z=8.5$
was adopted \citep{haardt96}.
Gas was allowed to radiatively cool \citep{sutherland93} and sink in the potential
well of DM halos. Whenever the gas exceeded a threshold
density of $n_{\rm H}=0.1{\rm cm^{-3}}$, five percent of it was turned into stars per
local free-fall time. Massive young stars exploded as supernova
after a time delay of 10 Myr corresponding to their average lifetime, and we modeled
these explosions using a Sedov blast wave solution \citep{dubois08}.
Cooling enhancement down to $10^4$ K due to the presence of metals is considered in the \mn simulation.
Potentially important physics such as AGN feedback, magnetic fields, and radiative
transfer were ignored in the simulation.
The simulation yields a sample of  3119, 6655, 11357, 15419, and 15999
halos at $z=6.1$, 5.0. 3.8, 2.5, and 1.5, respectively. This corresponds to all halos with  $ M_{\rm vir} \gtrsim 2.3 \times 10^{10} M_\odot$.}}

Let us first address the issue of the consistency of the orientation of large scale filaments
that pierce the virial sphere of dark halos\footnote{See Appendix~\ref{sec:geom} for a geometric analysis of the this effect}.
Figure~\ref{fig:correl-fil} displays the cross correlation of $n_{\rm HI}$ maps on the virial sphere,
 \rvir\,
for different pairs of redshifts as labeled on the figure.
This figure corresponds to the statistical average
over {\sc \small HEALPix} \citep{hivon05} maps of thresholded gas density measured at \rvir.
Since we are primarily interested in filaments, wall-like structures and void regions are excluded
by imposing a redshift-dependent threshold density. This is chosen as 6.6 times the
critical density of the universe, which
identifies the filamentary structures reasonably well at the resolution of the Mare Nostrum simulation.
The contribution from satellite galaxies is minimised by replacing the region
occupied by satellite galaxies with gas at the density threshold for star formation: $n_{\rm H}=0.1 {\rm cm}^{-3}$.
The cross-correlation coefficient plotted is  defined as
\begin{displaymath}
w(\theta)= \sum_{l,m} \langle a_{l m} b_{l m}^*
\rangle P_l(\cos(\theta))/\sqrt{ \sum_{l,m} \langle |a_{l m}|^2  \rangle
\sum_{l,m} \langle |b_{l m}|^2 \rangle }\,,
\end{displaymath}
where $a_{l m}$ and $b_{lm}$ are the harmonic transforms of the maps at the two
given redshifts, $P_l$ is the Legendre polynomial and $\langle \,\,\, \rangle$
stands for the statistical median over
halos within the considered mass range. 
The angle $\theta$ measures the separation between two pixels on the cross correlated maps, while the cutoff frequency is  $l_{\rm max} = 256
$.

Figure~\ref{fig:correl-fil} shows
that for halos with $M \sim 10^{10}-10^{11} M_\odot$
there is a significant correlation of
the directions of the gas infall between markedly different redshifts,
$w(0) \approx 0.3-0.4$ ($w(0) = 1$ implies total correlation). This is especially true at high redshifts.
Since halos of this mass have accumulated the bulk of their matter
over $\Delta z \approx 1.2 $ by redshift $z \approx 3.5$,
this value yields a {\it lower} bound
on the amount of temporal coherence achieved over finer time slices taken through
the epoch of assembly of the outer regions of a galactic patch.

{Let us now turn to the impact such a temporal coherence has on the advection of
gas angular momentum through the virial sphere.
As  the accretion-weighted specific angular momentum at virial radius reads:
\begin{equation}
\left<j\right>_{v_r} = \left| \frac{\sum_i m_i v_r \Theta(-v_r) \vec{\mathbf{r}}_i \times \vec{\mathbf{v}}_i}{\sum_i m_i v_r  \Theta(-v_r) } \right|,
\end{equation}
where $\Theta$ is the Heaviside step function, we measure it
by adding the contribution of the infalling\footnote{note that multiple accounts of backslash is statistically subdominant.} ($v_r<0$)  gas cells or particles within a shell with $0.95 \leq r/R_{\rm vir} \leq 1.05$.
Note that $\vec{\mathbf{v}}_i$ and $\vec{\mathbf{r}}_i$ are measured relative to the center of the halo and $m_i$ is the mass of cell $i$.
The values we quote are averaged over the whole halo sample at each redshift.}

 Figure \ref{fig:jvr_M}, {\sl left panel}, shows that the advected angular momentum {\em modulus} of the gas is
increasing with cosmic time and halo mass according to a trend
qualitatively consistent with that expected in the spherical collapse picture \citep{quinn92}.
The right panel represents the spin parameter, ($\lambda=j /\sqrt{2} R_{\rm vir} V_{\rm c}$,
\citealt{bullock01}), for the same halos. Note that, in this definition, the virial radius, $ R_{\rm vir}$, and the circular velocity, $ V_{\rm c}$,
are mostly set by the dark matter component. A residual trend in the evolution of gas spin as a function of redshift is clearly visible, along
with larger values of $\lambda$ gas compared to $\lambda$ DM. Note that, strictly speaking, $\left<\lambda \right>_{v_r}$ is plotted for the gas 
in contrast to $\lambda$ for the DM.  However, $\left<\lambda \right>_{v_r} ~ \lambda$ for the gas component in simulations where radiative cooling 
is accounted for (see \citealt{kimm11} for detail) which means that the comparison is valid. This does not mean that the spin of the freshly
accreted DM differs from that of the gas (it actually is very similar as shown in \citealt{kimm11}), but that the {\em total} spin of the dark matter halo 
does (see also \citealt{chen03, sharma05, brook11}).  Such an increase of gas angular momentum as a function of cosmic time was also noticed by \citet{brooks09}.

Figure~\ref{fig:jz_M}  displays the
ratio of the z-component of the advected spin of the gas  (where the z-axis is
chosen aligned with the spin of the dark halo) to the modulus of that spin.
The dense component of the advected (filamentary) flow, {\sl solid line}, is
distinguished from the diffuse material {\sl dashed line} using the following density
thresholds: $\times10^{-3}$, $5\times10^{-4}$, and $8\times10^{-5}$ $n_{\rm H}$ cm$^{-3}$ for
$z=6.1$, 3.5  and 1.8 respectively.
From the plot, it is clear that the dense infalling gas has a spin fairly well aligned with that of the
dark matter whatever the mass or the redshift considered and that in any case, this denser gas arrives much
more aligned than the diffuse gas at the virial radius.
Note that the z-component of the {\sl diffuse}  gas increases with redshift.
Figure \ref{fig:example-jz}  shows examples of typical maps of $\lambda_z$
and gives a good idea of the alignment of material with large values of  $\lambda_z$ with the main
direction of the filaments. As a matter of fact, such maps are quite representative of the
statistical result presented in  Figure~\ref{fig:jz_M}.

Finally, Figure~\ref{fig:cosinePDF}
represents the Probability Distribution Function (PDF) of the cosine of the relative angle between the advected angular momentum of the gas at different redshifts,
as labeled on the plot. It quantifies the correlation of the orientation of the momentum of the infalling gas as a function of cosmic time.
if there was no change in the direction of the angular momentum vector of a filament between redshifts, this PDF would be 
a Dirac distribution centred on $\cos(\theta)=1$.
On the other hand, if the direction of the angular momentum vector of a filament  
was completely uncorrelated between redshifts,  the PDF would be a constant independent of $\cos(\theta)$. 
The fact that it is markedly peaked around $\cos(\theta) =1$ shows 
that the amount of shifting/twisting/merging of filaments during cosmic time intervals is quite limited.
This figure complements the correlation found in Figures~\ref{fig:visual2} and~\ref{fig:correl-fil} for the orientation of the filaments 
themselves and corresponds, in redshift space, to the spatial correlation of the relative orientation of angular momentum along the filaments  described in Appendix~\ref{sec:space}.

{Together, Figures  \ref{fig:correl-fil}, \ref{fig:jvr_M}, \ref{fig:jz_M} and \ref{fig:cosinePDF} allow us
to draw the following statistically robust conclusions:
for the range of redshifts ($z>1.5$) and mass considered and the corresponding sub-grid
physics implemented in the \mn simulation, cold gas is advected through
filaments at the virial radius with an increasing spin parameter along a
consistent direction. The corresponding advected angular momentum orientation is
significantly correlated in time and its orientation is consistent with that of
the dark matter halo's spin. These trends are further established in a companion work
\citep{kimm11} which exploits much (minimum a hundred times) better resolved
individual galaxies in the \nut\ suite of zoom simulations. 
In that paper, it is
also convincingly demonstrated  that 
 the gas within
$\rvir$ has {\it more} specific angular momentum than its DM counterpart.}
Let us now explain these statistical measurements at the virial radius through
the dynamics of the surrounding cosmic web.

\section{Gravitational collapse of galactic patch}
\label{sec:story}
\subsection{ The origin of the angular momentum}
It is now accepted that the origin of the angular momentum of galaxies is in the initial distribution
of the matter velocities in the patches from which galaxies are formed. The shape of a typical protogalactic
patch is neither spherically symmetric, nor bounded by equipotential surface,
which results in a total
(i.e. integrated over the patch) non-zero angular
momentum that grows during linear stage of structure formation due to torques on
the patch from the gravitational tidal
field   \citep{peebles69,doroshkevich70,white84}.  The angular momentum growth
$\propto a^2 \dot D(t)$, where $D(t)$ is the growing mode of gravitational
instability, extends to the nonlinear Zeldovich regime \citep{porciani02a,porciani02b}. At this stage, the
specific
angular momenta of baryons and dark matter are equal.
While the global properties of the following nonlinear collapse of the patch are
well  captured by the spherical model or,
with more precision, by the elliptical approximation of the Peak-Patch-Theory \citep{bond96},
the detailed distribution of matter and motions in the patch are much more complex.
At intermediate mildly nonlinear stages, the dark matter in the patch assembles into the hierarchical cosmic web.
The gas follows the potential formed by the dark matter but is subject to
pressure forces whose balance
determines the level at which gas tracks the dark matter.
Numerical simulations (see Figure~\ref{fig:visual})
show that the cold
($T \sim 10^4 K$) gas forms rather narrow and smooth filaments that follow
the less well defined, wider and clumpier dark matter filamentary
overdensities.

\begin{figure*}
\centering
  \includegraphics[width=16.5cm,angle=0]{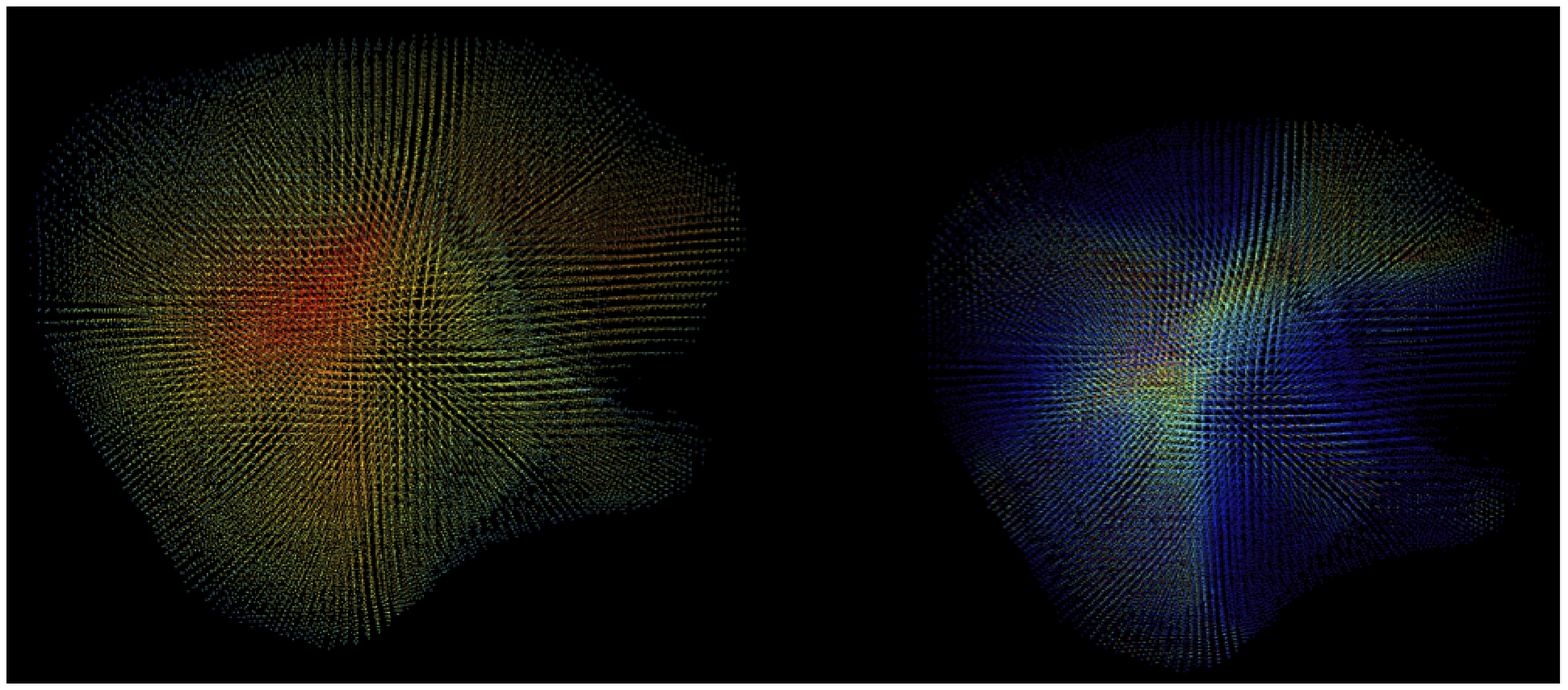} \hskip 0.75cm
  \includegraphics[width=16.5cm,angle=0]{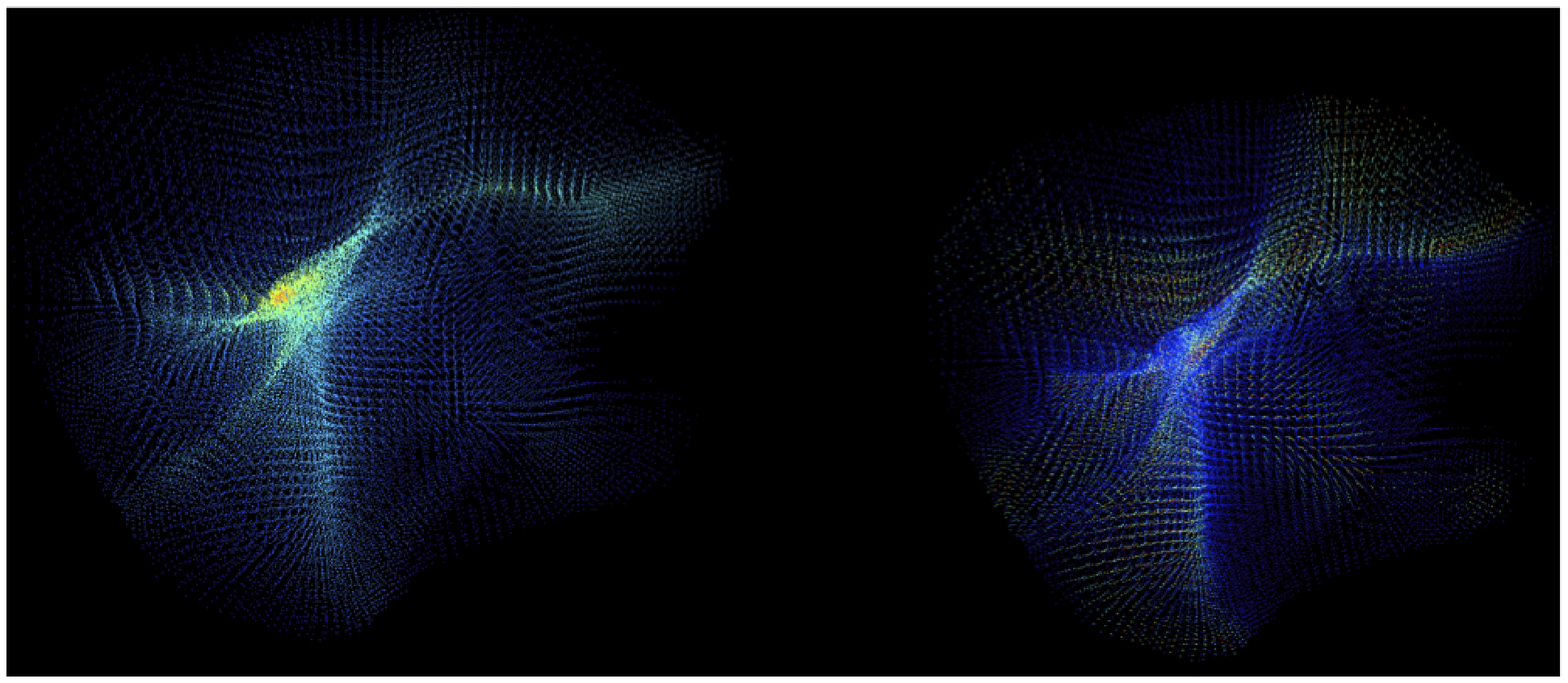}
   \caption{Colour coded density ({\sl left column}) and modulus angular momentum ({\sl right
column}) of the patch at early, $z > z_{\rm in}$, time ({\sl top row}) and at the
redshift
of the inner halo collapse $z=z_{\rm in}$ ({\sl bottom row}).
This figure illustrates the structure of the angular momentum distribution and
its later accretion in a collapsing
(proto-cluster) patch with a radius of 4.7 Mpc/h. The progenitors of the filamentary structures in the
initial conditions, i.e. the  central halo, $\sim R_{\rm smooth}=1.5 {\rm Mpc}/h$ in size, and the
filaments emanating from it,  are initially momentum poor,  since they are the regions
where velocity is predominantly radial. The high momentum particles (in blue) are in the
voids outside (not so much at the centres of voids but closer to
walls/filaments) where the velocities are more transverse w.r.t. the filaments.
At the linear stage the structures remain unmodified, with amplitudes of
velocities and, hence, angular momenta of the particles growing, but
displacement of the matter negligible. When the patch enters its non-linear
dynamical stage, the high momentum matter starts accumulating in the filaments,
making the filaments the locus of momentum rich material. Filaments change from
being momentum poor to momentum rich during the time needed for the matter to
travel from voids to filaments, which is similar to the time the inner halo
takes to collapse.  We observe that at redshift $z_{\rm in}$, exactly when the inner
halo virialises, filaments have changed to their momentum rich status.
This picture also shows that it takes time $t > t_{\rm in}$ for
the momentum rich  particles from the outskirts to reach the halo, since the
velocity of particles parallel to the filament is statistically lower. Filaments
provides momentum rich material at the accretion stage following initial
halo formation.
}
   \label{fig:elliptical-patch}
\end{figure*}

In this paper we are concerned with understanding the later time infall of
momentum rich material on an early formed proto-halo.
The deviation of the matter motion from semi-radial is especially significant in
the outer layers of the patch.
Let us consider a smoothed picture of the patch, with a smoothing scale, $R_{\rm smooth}$
selecting the inner halo region that collapsed at high redshift $z=z_{\rm in}$.
An example of such smoothed patch is shown in Figure~\ref{fig:elliptical-patch}
( see section~\ref{sec:3dexp}
 below for details about this experiment).
The exterior
region contains the mass that will be accreted during the further evolution to
redshift $z_{\rm col} < z_{\rm in}$ by which the whole patch collapses.

\begin{figure*}
\includegraphics[height=8.5cm]{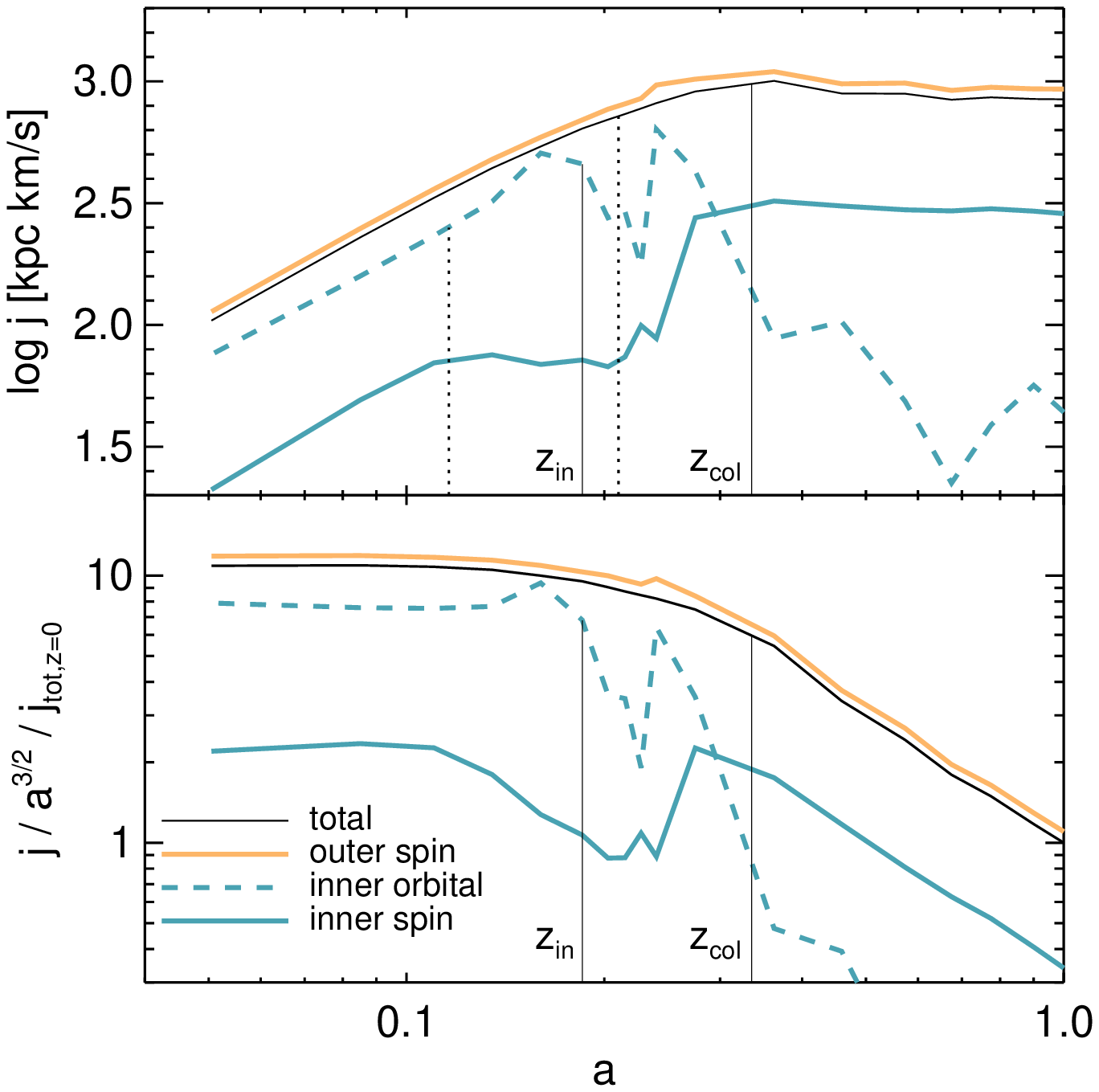}
\includegraphics[height=8.5cm]{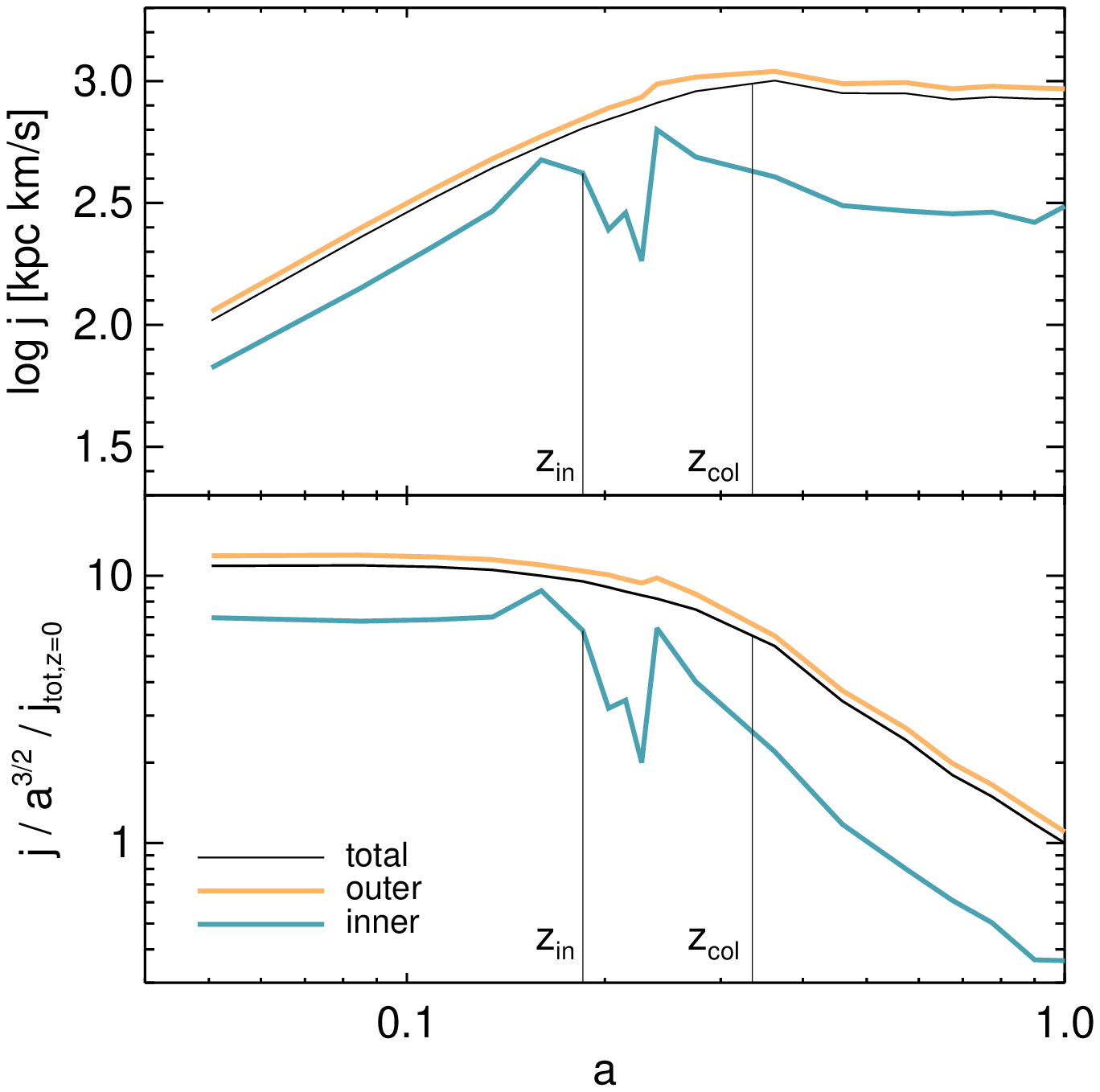}
\caption{Evolution of the specific angular momentum in a patch of 
$M = 4 \times 10^{11} M_\odot$, which is split into an inner core with 
$M_{\rm in} = 5 \times 10^{10} M_{\odot}$ and $R_{\rm inner} \approx
R_{\rm patch}/2$ and the outer envelope which contains the remaining 
7/8 of the total mass.
In the spherical top-hat model approximation, the patch collapses at 
around $z \approx 2$ which
we identify with $z_{\rm col}$. 
The inner core is found to correspond to an independent virialised halo at $z \approx 4.4$,
which is identified with $z_{\rm in}$. These redshifts are indicated by
vertical solid lines on the figure.  In addition, vertical dotted lines show 
the time of turnaround of the inner core at $z \approx 7.7$ and of the total patch at
$z \approx 3.8$. It is not accidental that the outer regions turns around
becoming nonlinear at approximately the same time as the inner core collapses, as it  
follows from our selection of the inner core to have roughly half the radius
of the whole patch.
The {\sl upper left} plot shows the evolution of the specific angular momentum
of the inner, outer and all the particles w.r.t. to the centres of mass (CoM) of 
the corresponding particle sets, i.e the spin of the total patch ({\sl thin black line}),
the spin of a cloud of particles in the outer region ({\sl yellow line}) and
the spin of the inner halo ({\sl cyan line}) respectively. In addition the orbital momentum
of the inner halo w.r.t. to the total CoM is shown as a {\sl dashed cyan line}.
The {\sl bottom left plot} shows the same data but divided by $a^{3/2}$, the expected
scaling of the angular momentum before turnaround in TTT.
We observe that the inner halo spin follows the expected behaviour, growing
$\propto a^{3/2}$ until turnaround, and then remaining constant
during its virialisation process.  At the same time the inner particles have
a significant orbital momentum w.r.t. the total CoM, which continue to grow
$\propto a^{3/2}$, essentially accounting for a  Zeldovich-like motion of
the inner core CoM relative to the total CoM. The situation changes when the outer 
envelope turns around: the orbital momentum of the inner halo is progressively converted
into spin as the CoMs of the inner and large patches are brought together. By the
time the whole patch collapses, the inner spin 
settles at the level that is enhanced, but
is still below the spin of the matter located in the outer parts of the patch.
The {\sl right panels} are similar to the left ones but
summarise the behaviour of the angular momentum w.r.t. the whole patch CoM only.
It again highlights the importance of the $z_{\rm in}$ to $z_{\rm col}$ time interval
for the distribution of the angular momentum in a forming galaxy.}
\label{fig:Jttt}
\end{figure*}

The dominant features of the geometry of the outer mass distribution are
filamentary bridges that emanate from
the central halo and extend to the neighbouring peaks. Saddle points along the
filaments mark the boundary of our patch.
Calculations of the peak-saddle cross-correlations in the Gaussian initial field
(Pichon et al. {\sl in prep}.) show that
the typical separation between a peak and the boundary saddle is $ \approx 2
R_{\rm smooth}$, with little sensitivity to the power spectrum slope.
Hence, our choice of the inner halo scale $R_{\rm inner}=R_{\rm smooth}$
to be half the radius of the patch $R_{\rm patch}=2 R_{\rm smooth}$,
allows us to just  resolve the filamentary environment of the proto-halo.
We relate our picture to the tidal torque theory
\footnote{ But, in contrast to the classical tidal torque theory (TTT)
framework, we investigate the properties of our patch in greater detail,
effectively using a filtering scale roughly twice as small as what is used in
TTT.  This constitutes departure from the quadratic representation of the
density and tidal fields
that,  we argue, is important in driving the anisotropy of the delayed infall.}
in Figure~\ref{fig:Jttt} where the evolution of the angular momentum of 
dark matter is measured separately for the inner core and the outer envelope of
a representative Milky Way-like patch (the \nut\ simulations see \citealt{kimm11} for details).

While the dynamics of the smoothed inner halo can be approximated by the
ellipsoidal collapse with matter infalling along mostly radial
trajectories, the filaments represent structures in which the outer matter
with significant transverse velocities accumulates. For this matter the velocity
component parallel to the filaments is statistically reduced
so the drain of matter from filaments dominates the accretion onto the inner
halo at the later times. Note that the time it takes for the inner halo to
collapse is similar to the time for the outer filaments of the same scale to
turn around \protect\citep{1998wfsc.conf...61P}, reaching overdensities
$\delta \sim 10$ and forming potential troughs to channel cold gas.

\begin{figure*}
\includegraphics[height=8cm]{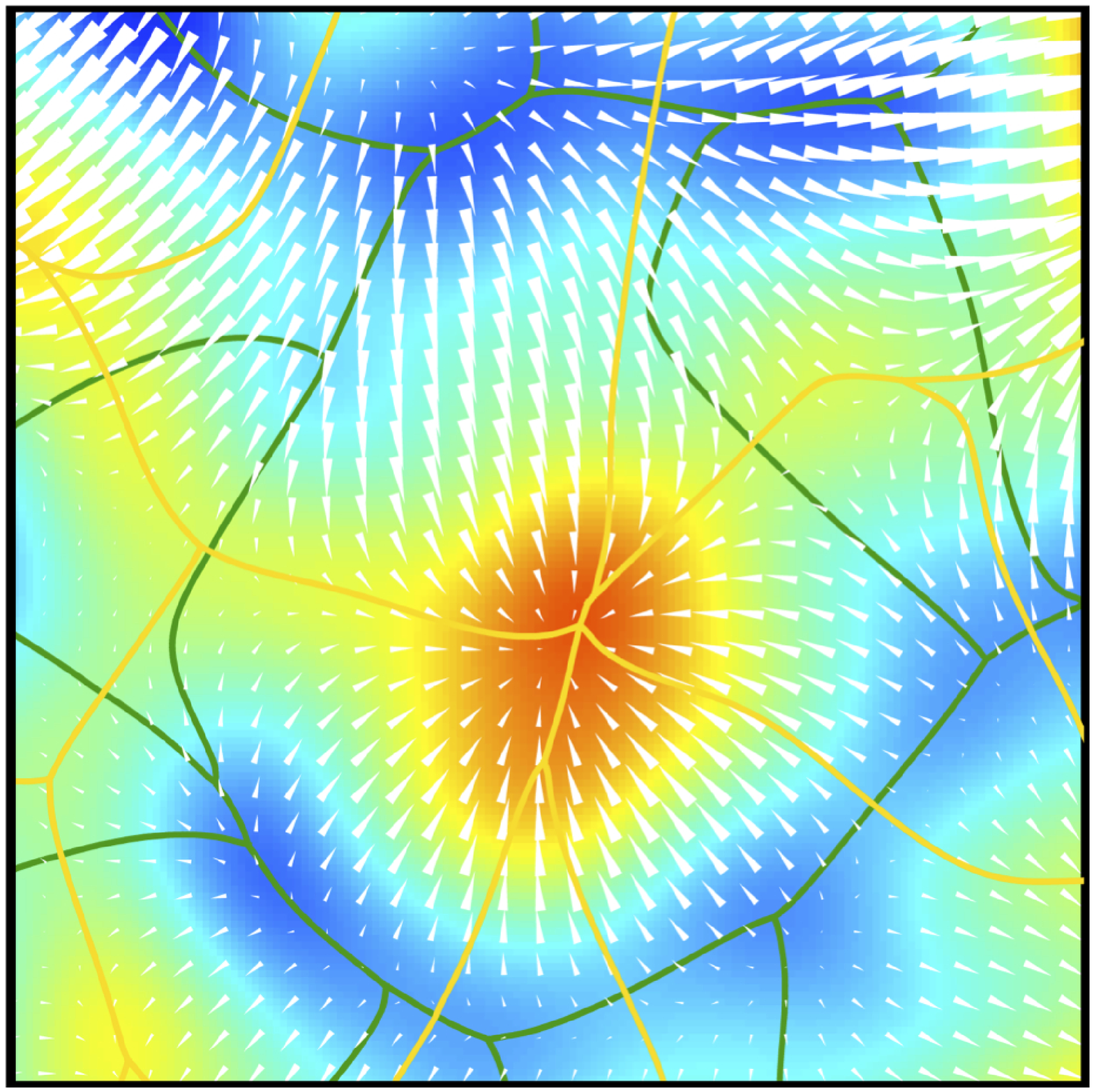}
\includegraphics[height=8cm]{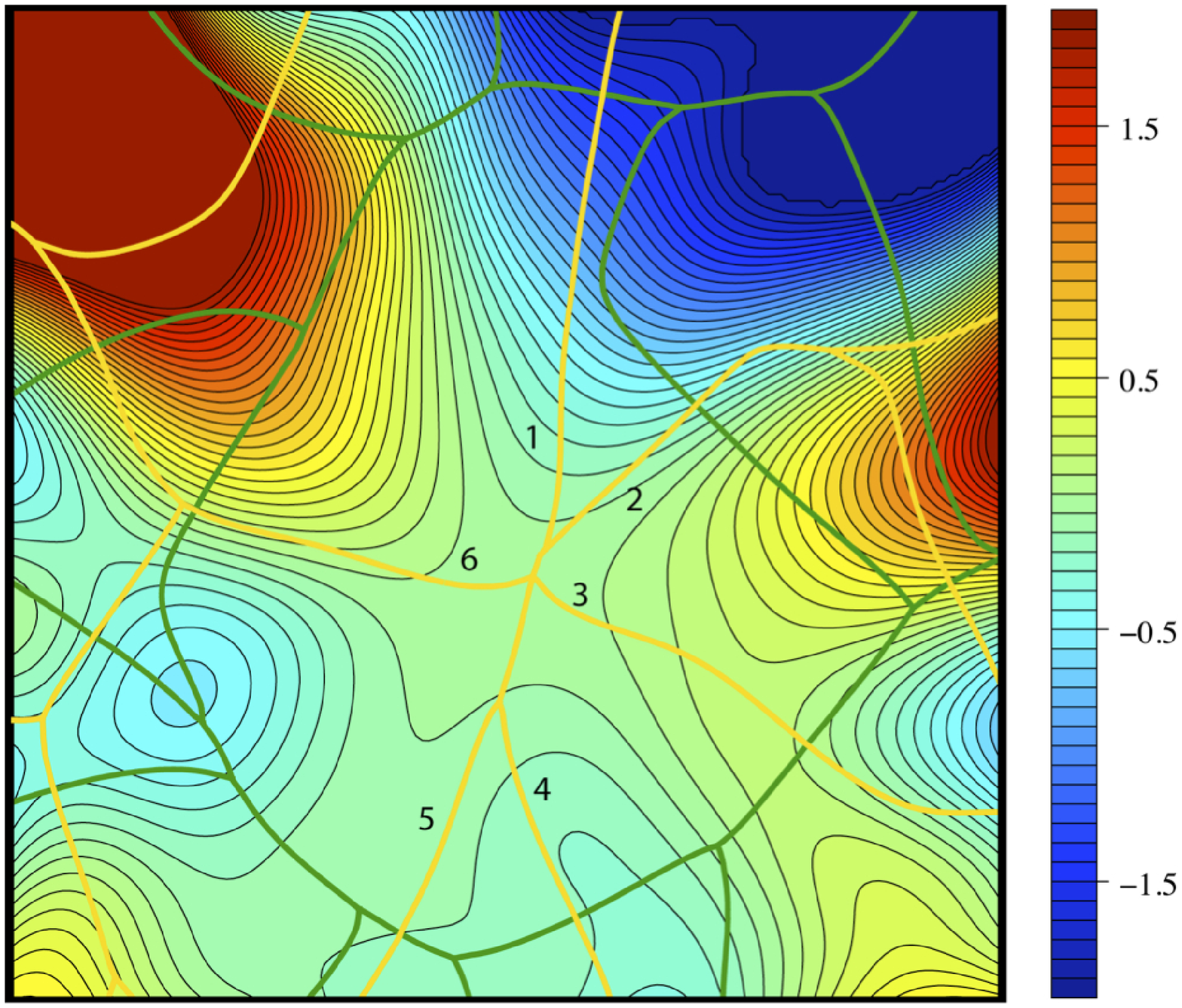}
\caption{{\sl left panel:}  2D density field and 2D
initial velocity field measured w.r.t. to the central peak.
{\sl Right panel:} the corresponding angular momentum w.r.t. the central peak position.
The green colour corresponds to small angular momentum, as, for example, near
the central peak. Both deep blue and red colours are regions of high magnitude
angular momenta, of opposite signs, clockwise and anti-clockwise, respectively.
In both panels the skeleton of the progenitors of the filamentary
structure is superimposed as gold lines. Green lines map the boundary of the
patch according to the density gradient flow.
The most massive filaments (\#2 and \#6) to be formed are those that form
overdense
bridges to the neighbouring halos. These filaments are indeed
the loci of the divergence of the velocity field and are the progenitors of
long-lived structures along which the matter  drains onto the central halo
with  a time delay relative to the isotropic infall. They define
the directions along which much of the outer material will reach the halo.
(a note of caution: one should not consider the initial velocity field as
delineating flow lines for the future infall of material).
In the example given,
these directions are not generally coincident with the shape of the central
elliptical proto-halo, but are defined by the position of the neighbouring halos and voids
and the resulting velocity field structure. However, it is when there is 
an alignment of the peak and the shear of the velocity field
that the most prominent filaments arise \protect\citep{bkp96}.
These dense filaments provide  deep enough
potentials to contain the $T \sim 10^4 K$ gas.
The other (e.g., \#1,\#3,\#4 and \#5) filaments of the skeleton, although
mapping critical lines of the density gradient by construction, may not be
associated with particularly notable overdensities. They reflect short-lived
anisotropies of the collapse of the halo, as, for example, might arise from a
mismatch of the orientation of the original peak with the velocity flow.
}\label{fig:2Dpatch}
\end{figure*}

For illustration,
let us first observe the typical {\sl initial} properties of a patch around the
peak in 2D, as shown in Figure~\ref{fig:2Dpatch}.
The theory of the skeleton of the cosmic web \citep{novikov06,pogo09} provides
an
approximate boundary of the collapsing patch by defining it as
surfaces that pass through the saddle points on the filaments linked to the
central peak. This partitioning is not exact in view of the future non-linear
evolution of the structure, but is closely related to the partitioning of the
initial velocity flow pointing to the peak. Importantly, it illustrates the
generally non-spherical nature of the initial proto-halo.

The map of the angular momentum in Figure~\ref{fig:2Dpatch} shows that at the initial stage both the central
region and primary dense filaments (the filaments that form bridges to
neighbouring haloes, in our example tracked by  branches $2$ and
$6$ of the skeleton) typically contain matter with initially little angular
momentum. Particles with high angular momentum are spread outside the
dense structures in regions with alternating sense of rotation. As the
evolution of the patch progresses, these filaments will collect the high angular
momentum
from the outside material and, being responsible for  the late accretion towards the central peak, supply angular momentum after the central halo forms.
The angular momentum in a filament is dynamically contained in the displacement of its
apex w.r.t. the center of the mass of the halo, and its residual transverse motion.
Both are generic features of filamentary structures formed in gravitational
collapse, whose relative impact may vary with the mass of the filament and
the structure of the patch. Note finally that the amount of angular momentum cancellation
expected during shells crossing is bound to be less important in 3D than for this
oversimplified setting.

\subsection{The 3D dynamics of the collapsing patch }
\label{sec:3dexp}

\begin{figure}
\includegraphics[width=8.5cm]{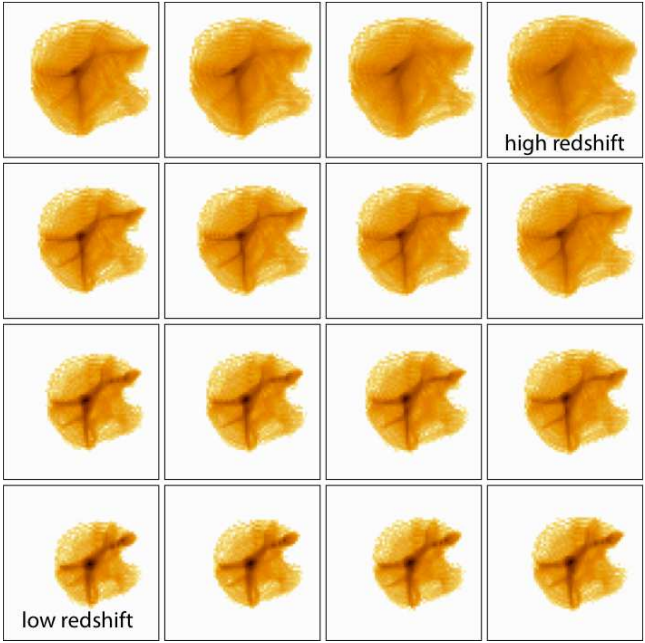}
\caption{Projected view of the dark matter density in a collapsing patch with smoothed out ICs as a function of cosmic time {\sl from top to bottom and right to left.}
 As these maps reflect the entire merging history of a given halo, they clearly show that the incoming direction of the infall  remains fairly constant with time,
 with a residual drift, which reflects the advected angular momentum along the filament (see also Figures \ref{fig:visual2} and \ref{fig:skel}).}
\label{fig:projection}
\end{figure}

The fate of the total angular momentum within the patch is encoded
within the initial condition and the later evolution of the surrounding tidal
field \citep{doroshkevich70}. As a larger and larger
fraction of the patch turns around, it becomes insensitive to the latter, hence
its enclosed angular momentum is conserved.  However, since the
Vlasov-Euler-Poisson set of equations is non-linear,  the detailed timing
and the geometry of the flow matters in determining how and where this
angular momentum is redistributed locally. 
In the first stage of the gravitational collapse, angular momentum is mostly
in the transverse motion within the expanding voids, then its partially cancels out in walls as the void shells cross; it cancels out again in filaments,
and again towards the center along filaments.
In effect, the measured flux of momenta at various radii all depends i) on the
subtle imbalance in transverse motion there was to start with (i.e. on the
relative geometry of surrounding voids), ii) on the relative mass in these
different structures, iii) on the external and internal torquing between these
components, and finally iv) on the fraction lost to spinning up substructures
forming along.
It is therefore not straightforward to visualize  the different stages of the
transport in three dimensions.

To demonstrate in 3D the development of the filamentary
structure in the outer regions of a collapsing patch and its role for angular 
momentum transport,
let us carry out the following idealised numerical experiment.
We generate the initial conditions of a $100 h^{-1}$Mpc  $\Lambda$-CDM
simulation  using  {\texttt{MPgrafic}} \citep{prunet08}  with $256^3$ dark matter particles.
The {\sl initial} density and velocity cubes are then smoothed with a Gaussian
filter of  $\sigma=1.5$ Mpc$/h$. 
The smoothed IC simulation is then run down to $z=0$ where a friend of friend  (FOF) catalog is constructed.
Within that catalog we choose a somewhat massive halo ($\sim 2 \times 10^14 M_\odot$),
define the halo patch as the Lagrangian extension of all particles which end up
within the FOF halo at $z=0$, and follow the evolution of these particles. 
The patch therefore consists of an inner core that is defined by the smoothing length
and an outer envelope that develops filamentary structure which we focus upon.
We stress that the only density field which we ever smooth is the initial one.
Such a low resolution simulation allows a qualitative discussion of the dark matter patch structure.
Appropriate physical conditions at the correct scales and redshifts need to be considered when extending the scenario to
describe gaseous filaments. 

Figure~\ref{fig:elliptical-patch} shows the density 
structure of the patch as it evolves
from its initial conditions, and the distribution of angular momentum amongst the
particles in the patch.  While high angular momentum particles initially 
are found 
predominantly in the voids, outside the inner core and filament progenitors in the
initial conditions, they aggregate to the filamentary regions, as the latter
grow in density. Even with the cancelling effects
of transverse shell crossing taken into account, the filaments become
the locus of relatively high angular momentum by the time,
$z_{\rm in}$, of inner core collapse. The angular momentum is then contained
in the residual transverse motion of the filaments and
their misalignment with  respect to the centre-of-mass of the patch.

Figure \ref{fig:projection} displays a time sequence projection of the log-density
within that patch.  The filaments connecting the central halo to the edge of the
patch are clearly visible,  and their relative motion does reflect the tidal
field within that patch. In the first stage of the gravitational collapse, both
the central peak and the surrounding voids compete to, respectively,
attract and repel the dark matter   in the outskirts of the patch.
From a distance, this figure suggests that indeed the orientation of the filaments does not
 change much over the course of the collapse of that patch (see also Figure~\ref{fig:skel} below).

Figure \ref{fig:trace0}
\begin{figure*}
  \center{\includegraphics[width=8.25cm]{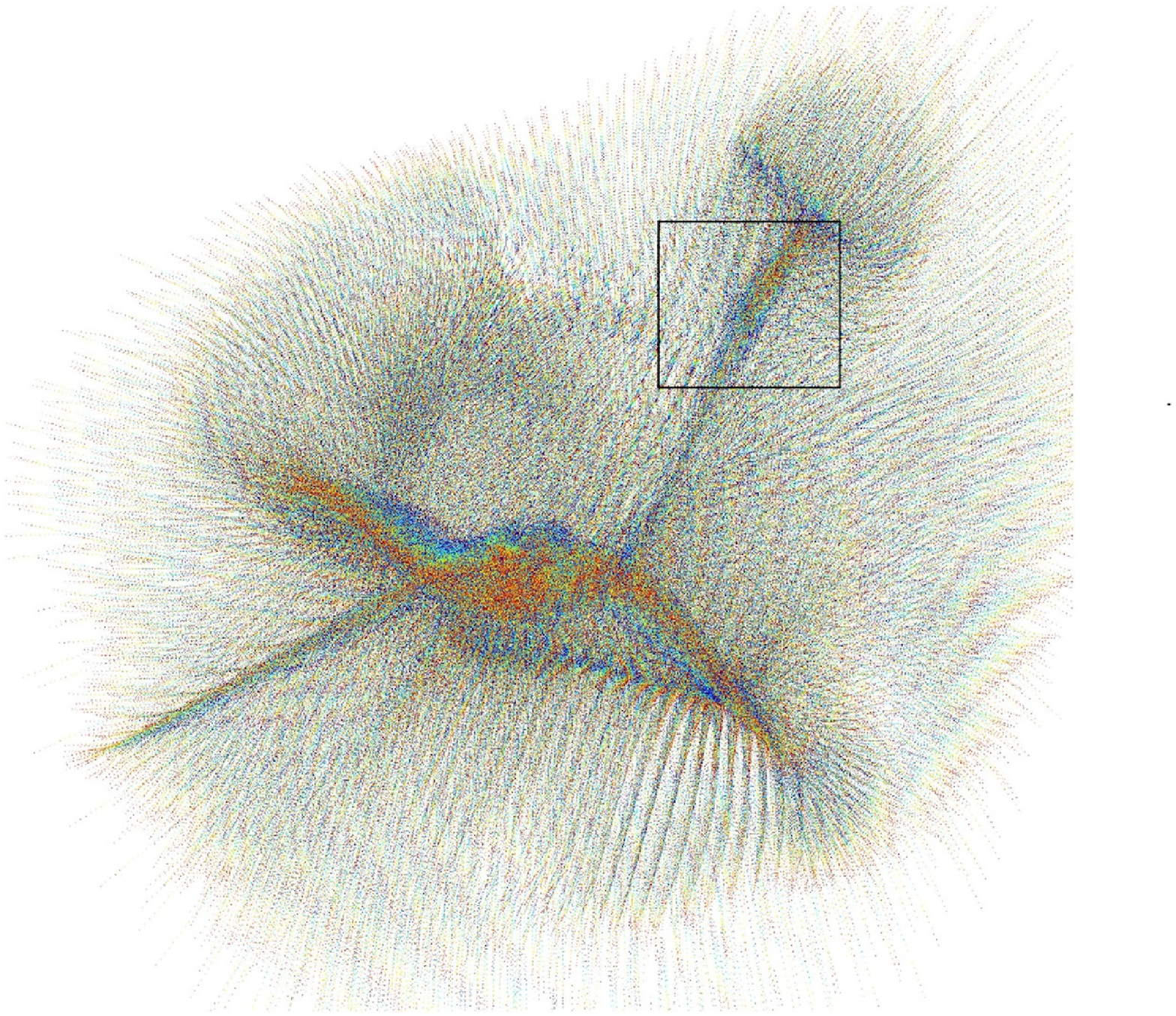}\hskip 0.5cm
\includegraphics[width=7.5cm]{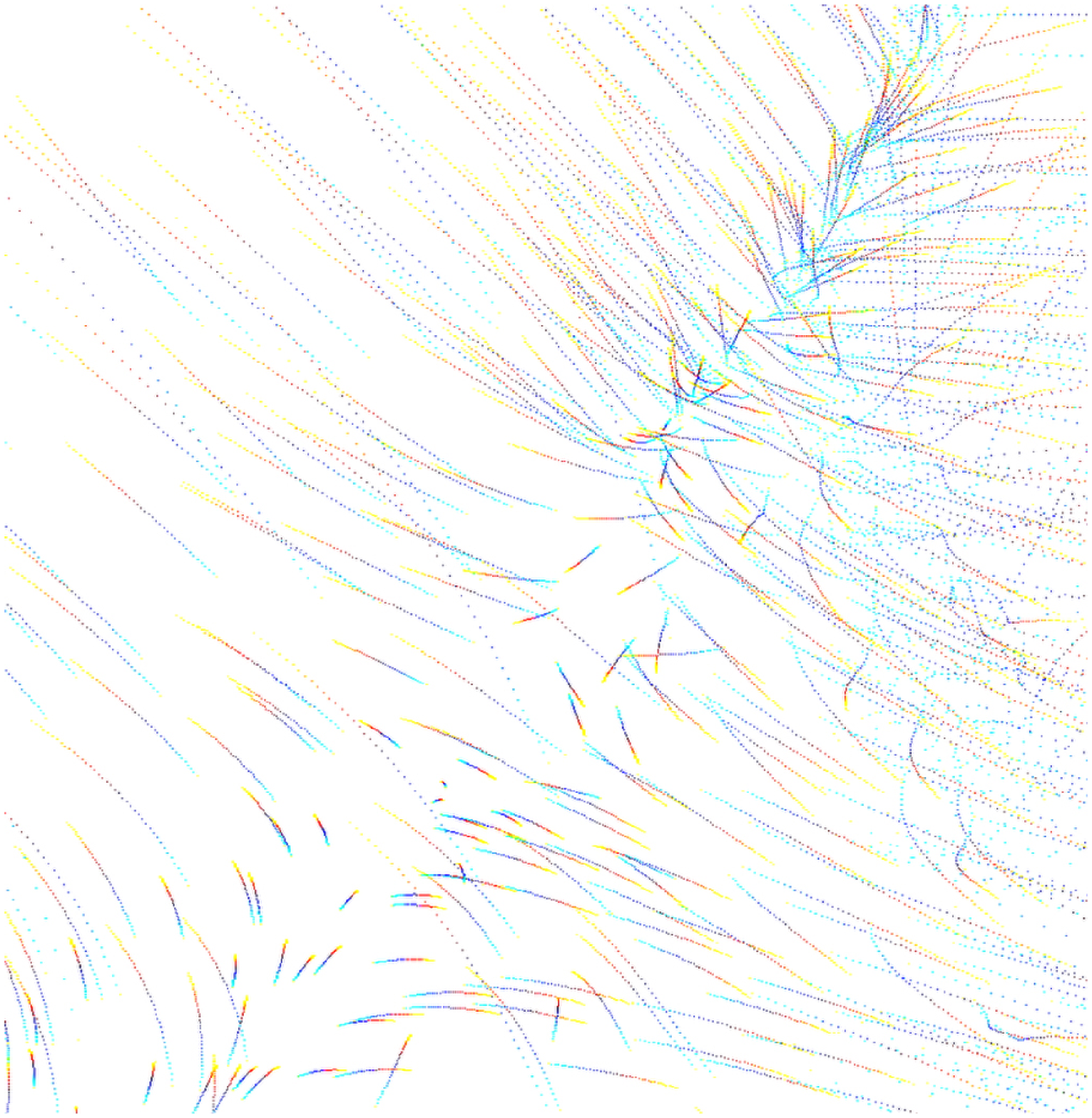}}
 \caption{
{\sl Left:} example of trajectories of dark matter particles, colour coded by
redshift from red (high redshift) to blue (low redshift). These trails first converge towards the filaments and then along the
filaments towards the central halo: their motion account for both the orbital
angular momentum of the filamentary flow and the spin of satellites formed within
those filaments. {\sl Right:} a zoom of the left panel corresponding to the shell
crossing between two walls leading to the formation of the north east filament.
Some of the angular momentum lost in the transverse motion is given to the spin of
structures forming within that filament; the rest is converted into the angular
momentum of the filament. The corresponding animation is available for download at {\em\tt http://www.iap.fr/users/pichon/rig/}.
\label{fig:trace0}}
\end{figure*}
shows the trajectories of particles colour-coded by redshift
over a range of cosmic time.
 Inspecting  these trails in 3D is instructive and led us to the scenario
presented in this paper. It appears quite clearly that the trajectories of the
dark matter particles
 initially within the voids of the patch present a sequence of inflections.
These inflections correspond to shells crossings, when the flow either reaches a
wall or a filament or finally  the central peak. For instance, the right panel
of Figure \ref{fig:trace0}, which corresponds to a zoom of the north-east
filament,
presents such whirling trajectories, where  a fraction of the transverse flow
within the surrounding walls coils up and is converted into a spinning, sinking
feature within the filament.  Through this process, a fraction of the orbital
angular momentum within the large scale structure is transformed into spin, while the
residual transverse motion is converted into the drift of  the filament.
\footnote{This cosmic flow within filaments is also consistent with the measured spin of dark halos in filaments (Codis et al. {\sl in prep.})}
Indeed,
Figure \ref{fig:skel} demonstrates this on the smoothed IC patch by tracing the
filaments of different snapshots using the skeleton \citep{sousbie11}, a code which basically identifies
the ridges connecting peaks and saddle points  of the density field.
Here the ``persistent skeleton`` was computed from the dark matter particles
within
the patch, and co-added for a range of redshift while keeping fixed the position
of the most bound particle. The residual distortion from one snapshot to
another reflects  the drift of the momenta rich filaments.
\begin{figure}
\includegraphics[width=8.5cm]{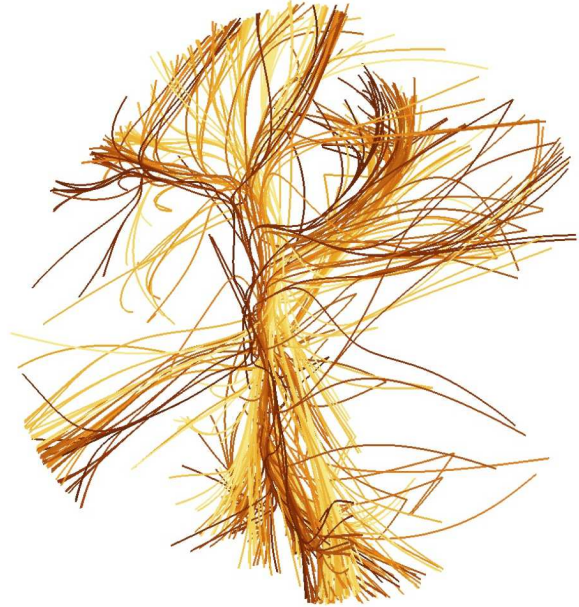}
\caption{The sweeping skeleton within the inner region of the patch shown
in Figure \ref{fig:projection} colour coded by redshift from dark (high $z$) to light (low $z$).
The central peak is in the middle of the figure.
To zeroth order, this figure suggests
that the direction of the filament does not change much.
Looking more closely, from one redshift to another, the various branches
of the skeleton do slide with cosmic time. Indeed, the net transverse motion
following shells crossing  of the filament will induce a residual drift of the
filament, which the skeleton captures. At the top of this plot, note the top left-most fork 
before a satellite which will merge with the central halo at a later time.}
\label{fig:skel}
\end{figure}

\begin{figure*}
\includegraphics[width=6.cm]{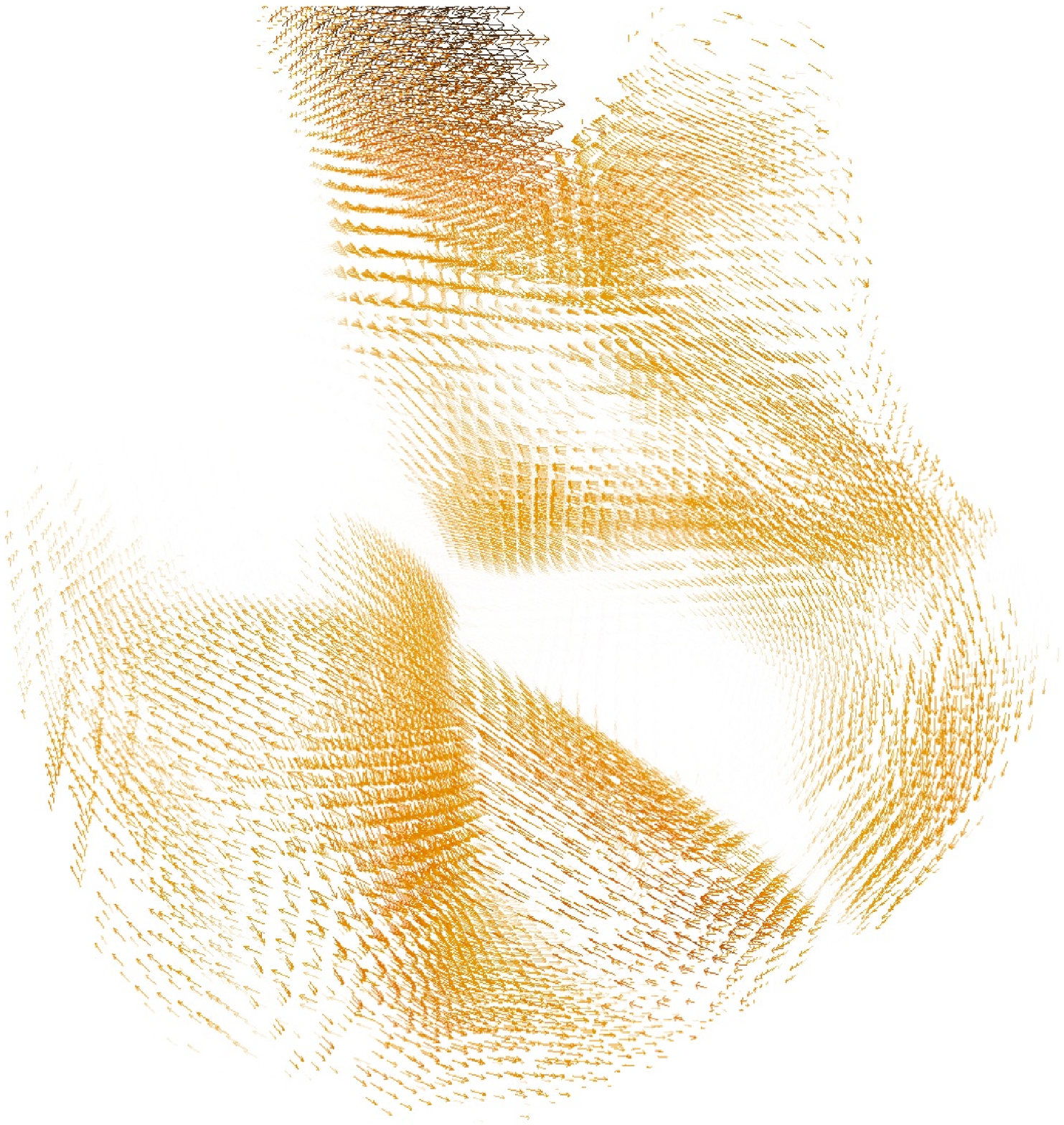}\includegraphics[width=6.cm]{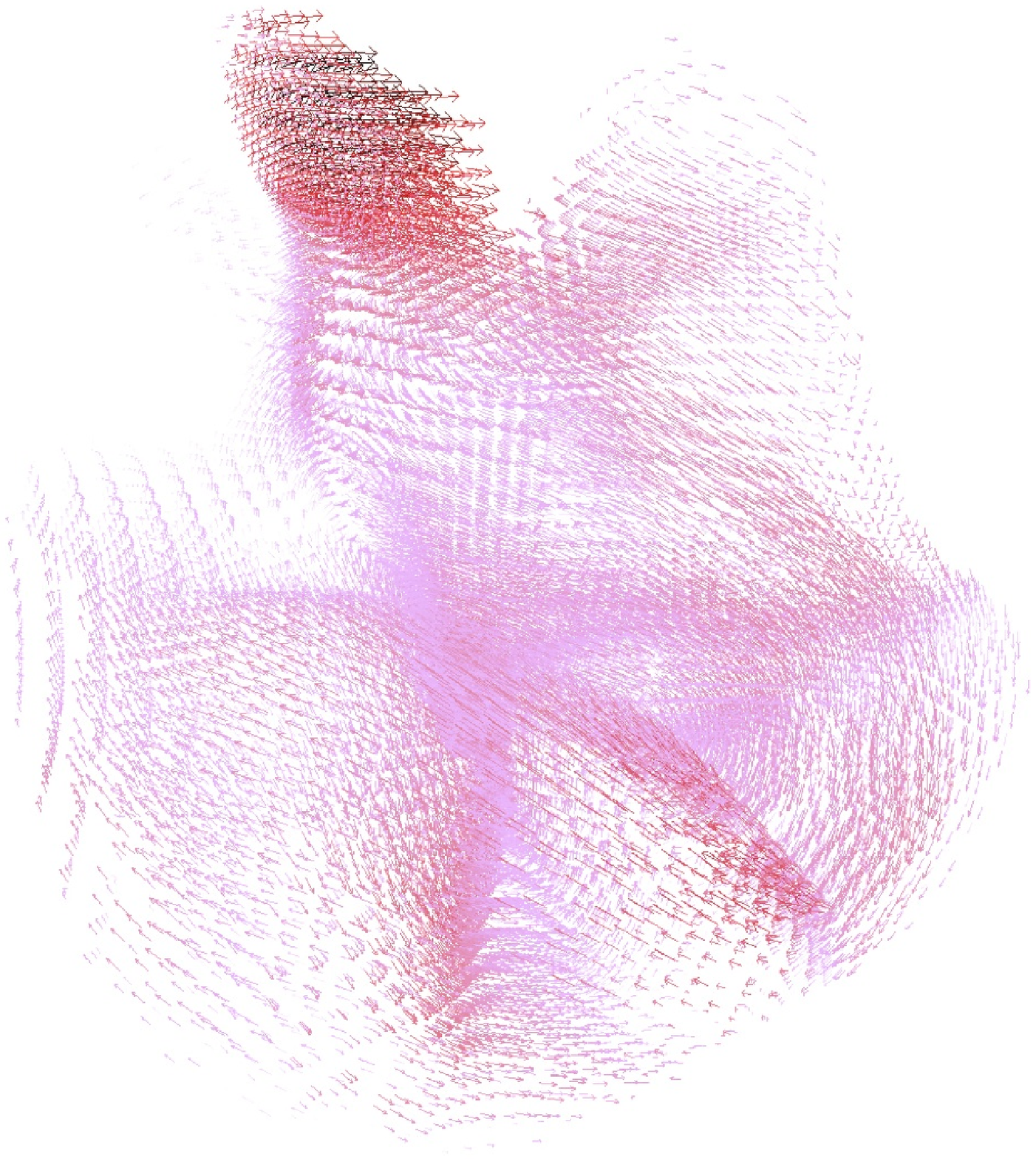}\includegraphics[width=6.cm]{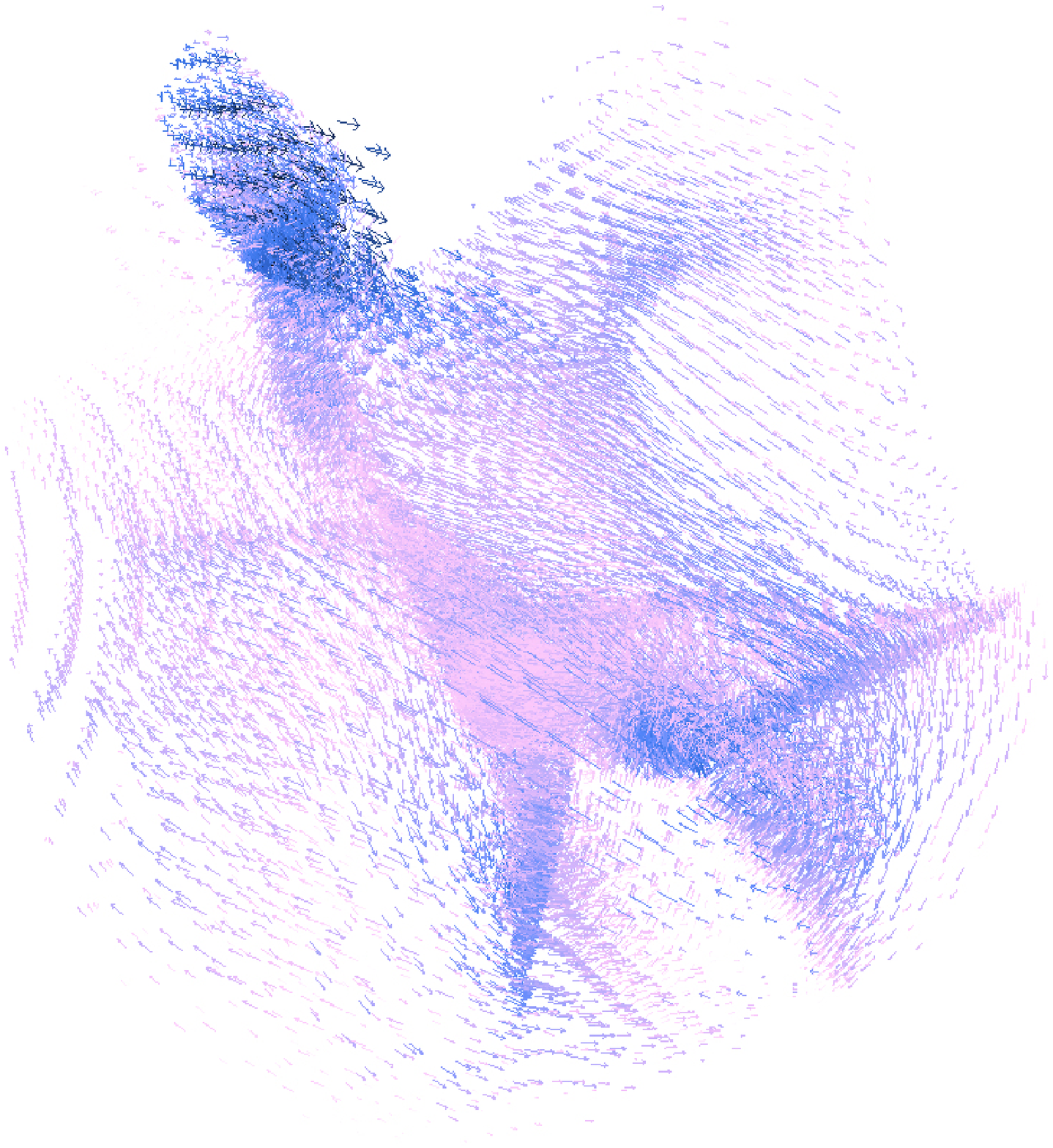}
\caption{Redshift sequence ( $z> z_{\rm in}$,  $z= z_{\rm in}$ and , $z<z_{\rm in}$) displaying the 3D angular momentum vector field, colour-coded by amplitude  (with the same dynamical range throughout; dark corresponds to large angular momentum, light to intermediate angular momentum; low angular momentum is not shown). Note that the choice of colour is arbitrary and is only dictated by the requirement that 
the colour be different for different redshift bins. {\sl The left panel} clearly shows that the voids are angular momentum rich, the middle panel shows that near $z_{\rm in}$ the filaments have partially advected (and partially canceled out) a fair fraction of the void's momenta, while the {\sl right panel} shows that the remaining angular momentum indeed is carried in through the filaments.
Note that the region of high and intermediate momenta do not overlap much between  $z> z_{\rm in}$ and  $z<z_{\rm in}$, as expected given the flow. The corresponding density is displayed in figure 7 and  10 along different orientations.
A rough estimate gives that
 filaments, walls and voids  contain 60 \%,  30 \%,  10 \%  of the specific angular momentum at $z=z_{\rm in}$.  The corresponding animation is available online at {\em\tt http://www.iap.fr/users/pichon/rig/}.
 }
\label{fig:momentum}
\end{figure*}

Let us now look at the actual angular momentum advected and canceled in the flow.
Figure \ref{fig:momentum} displays vector fields of the momenta for three
snapshots, roughly corresponding to times when (left panel) most of the momentum
still lies in the voids, (middle panel) a significant fraction had started
migrating in the filaments and walls and (right panel) most of the flow had
converged into the central object and filaments. On this figure, the colour
coding reflects the amplitude of the angular momentum and spans the same range of values
across the three panels. It follows that the amplitude of the advected angular momentum
is lower along the filaments than in the voids (owing to the cancellation while
it is being formed),
but displays a significant gradient along the north and the west filament on the
right panel.
Of course these idealised numerical experiments raise the so called  ``cloud in cell problem": here we focused on a ``hot dark matter'' picture of things where competing processes on smaller scales do not disrupt the larger scale structure of the flow. This assumption is reasonable, as we expect the larger scale distribution of matter to impact most the later angular momentum advection.
The hydrodynamical results of Section~\ref{sec:hydro} suggest they are not statistically dominant, at least beyond
redshift \footnote{ Substructures on smaller scales may become relevant at lower redshift when dark energy kicks in and dries out the gas supply, for a subset of massive halos embedding elliptical galaxies
 (see  \citealt{tenreiro11} and the discussion below).} $z=1.5$.

\section{Discussion}
\label{sec:conclusion}
This paper has presented a series of results concerning the nature of the angular momentum rich
dynamical gas flows at the virial radius of collapsed dark matter halos. Measurements were carried out using the
AMR Mare Nostrum simulation  \citep{ocvirk08}  and allowed us to draw the
following conclusions: at redshift 1.5 and above, gas inflow through the virial
sphere is focused (preferred inflow direction) consistent (orientation of advected angular momentum steady in time) and amplified
(increasing amplitude of advected angular momentum as time goes by).
The qualitative analysis of very simple 2D initial conditions and idealised
dark matter  simulations have allowed us to explain this coherent flow in terms of
the dynamics of the corresponding gravitational patch.

\subsection{Advection of angular momentum along cold flows}
In view of these findings we sketched the following scenario for the gas flow along cold streams.
 In the outskirts of a forming halo, large scale flows arise as the surrounding voids expel
gas and dark matter. As the flows coming from opposite directions meet, dark matter undergoes shell-crossing whilst the gas shocks, cools and collapses into
the newly created walls and filaments at the boundary between voids. Each of these boundaries acquires a net transverse velocity which reflects the asymmetry
between the voids it divides. Within the boundary, the longitudinal component of the flow then drags gas towards the growing halo, advecting angular momentum in the
process.  As the transverse velocity should be similar along the boundary, the later the infall onto the halo, the further it originates from, the larger the angular momentum
it brings (lever effect). Thus the orientation of the advected angular momentum at the virial radius of the halo is steady in time, as it is piloted by this large-scale transverse
motion which in effect is encoded in the initial conditions of the patch. It also means that gas hits the virial sphere of the halo along a preferred direction: that of
the incoming filaments/walls.

We are now in a position to try and answer the two questions raised in the introduction, namely "how and why is gas accreted onto the galactic disc?".
The answer to ``how'' is:  so as to build up the circumgalactic medium through the direct accretion of cold gas with ever increasing and  (fairly well) aligned
angular momentum. The answer to why is: because the internal dynamics of the cosmic web within
the peak patch produces such a coherent flow. Angular momentum rich inflow is delayed compared to radial inflow as a fraction of the
patch first flows away from voids
and into the walls and filaments. Only then is this material brought back in the direction
of the halo.

Hence the anisotropic distribution of angular momentum outside of the collapsed
halo is critical to explain the {\sl  formation} of thin galactic discs.
It means that the gas arrives at $\rvir$ in two distinct flavours: dense versus
diffuse, as outcomes of different dynamical histories within the peak patch. The
denser phase is produced as gas collapses into walls and filaments where it cools
and which further feed the central halo.
The diffuse phase is  either gas that is accreted directly onto the halo, or the hot gas
which was unable to cool due to low density, and was not confined to
filamentary structures.
As the result, the hot gas does not display a preferred alignment w.r.t. the
dark halo's
spin (Figure~\ref{fig:jz_M}), as this coherence is only achieved via
filamentary infall in which it did not participate.
Even though the mass
involved in filamentary flows can be small compared to the total gas mass
involved in the formation of the central galaxy at any given time, it should
play a critical role in supplying the circumgalactic region constructively with
increasingly angular momentum-rich gas,
while at the same time minimizing the destructive impact of incoming
substructures on the existing disc.
In short, in this scenario, discs are in fact {\sl produced by}, rather than {\sl shielded from} the cosmic environment.

\subsection{ The dynamics of the gas within $R_{\rm vir}$}
\label{sec:circum}
Though the dynamics of
the infalling gas within $R_{\rm vir}$ is complex and is the topic of companion
papers  (\citealt{kimm11};Tillson et al. {\sl in prep.}; Powell et al. {\sl in prep.}),  we qualitatively describe here what happens
to the cold gas brought by filamentary streams into $R_{\rm vir}$ as it reaches
the outskirts of the central disc (the circumgalactic medium). 
Simulations show  these filaments penetrate the $\sim R_{\rm vir}/10$ region on a infalling orbit, while roughly preserving
their radial velocity and width \citep{dekel09}, radiating away the kinetic energy acquired
during the free fall \citep{goerdt10}. They typically overshoot the centre of the halo until they encounter
counter-falling diffuse gas against they radiatively shock \citep{powell10} and resume their plunging orbit
on a much less radial orbit progressively spiralling in towards the
inner disc. 

A limited loss of the coherence inherited from the large scale
structure is expected to occur: as the filamentary inflow has a geometry which is steady in time,
so should the shock. Similarly the gravitational interaction
between infalling satellites and the existing triaxial halo and pre-existing
proto-disc should only have a time limited impact on the reshuffling of angular momentum
within that region.
In a nutshell, we expect the consistency of the cosmic flow to overcome, in the long run,
the occasionally messy circumgalactic environment of forming discs. 
Using the \nut\ \citep[see below]{powell10}  suite of highly resolved hydrodynamical simulations,
\cite{kimm11} find that the specific angular momentum of the gas is at least
twice as large as that of the dark matter in the DM halo
$R_{\rm vir}/10<r< R_{\rm vir}$ (see also e.g. \citealt{chen03, sharma05, stewart11}). 
More importantly they find that this specific angular momentum is, most of the time, well aligned 
with that of the central galactic disc region. This means that the outer halo region acts like a reservoir 
of rapidly spinning cold gas, which coherently builds the disc inside out. We therefore conjecture that, 
at least {\sl statistically}, disc galaxies owe the survival of their thin disc to the
consistency of the  cosmic infall.

\subsection{Linking the gas flow within a disc galaxy to its LSS filamentary origin above redshift 1.5}
\label{sec:NUTtrace}

\begin{figure*}
\includegraphics[width=5.75cm]{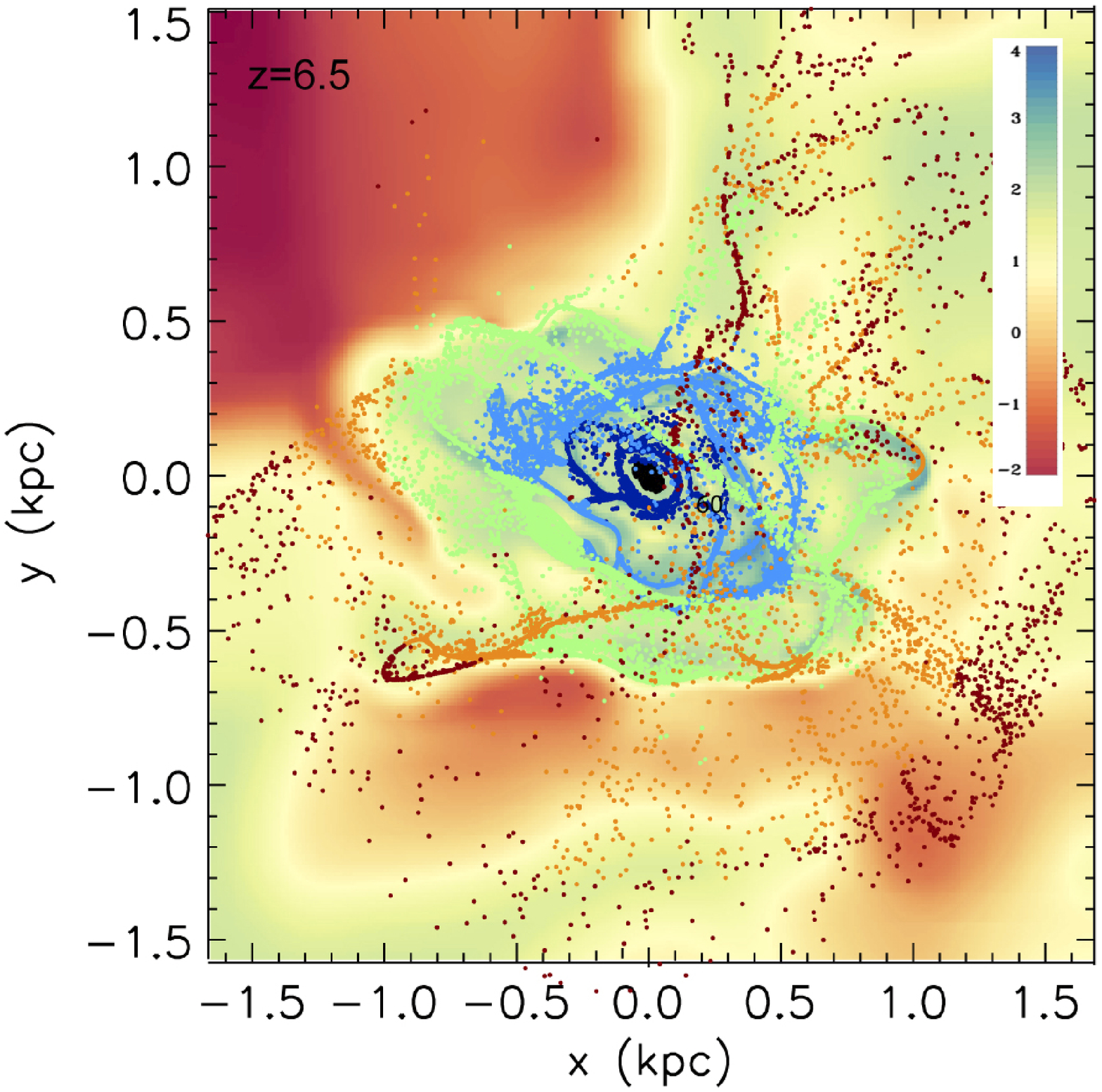}
\includegraphics[width=5.75cm]{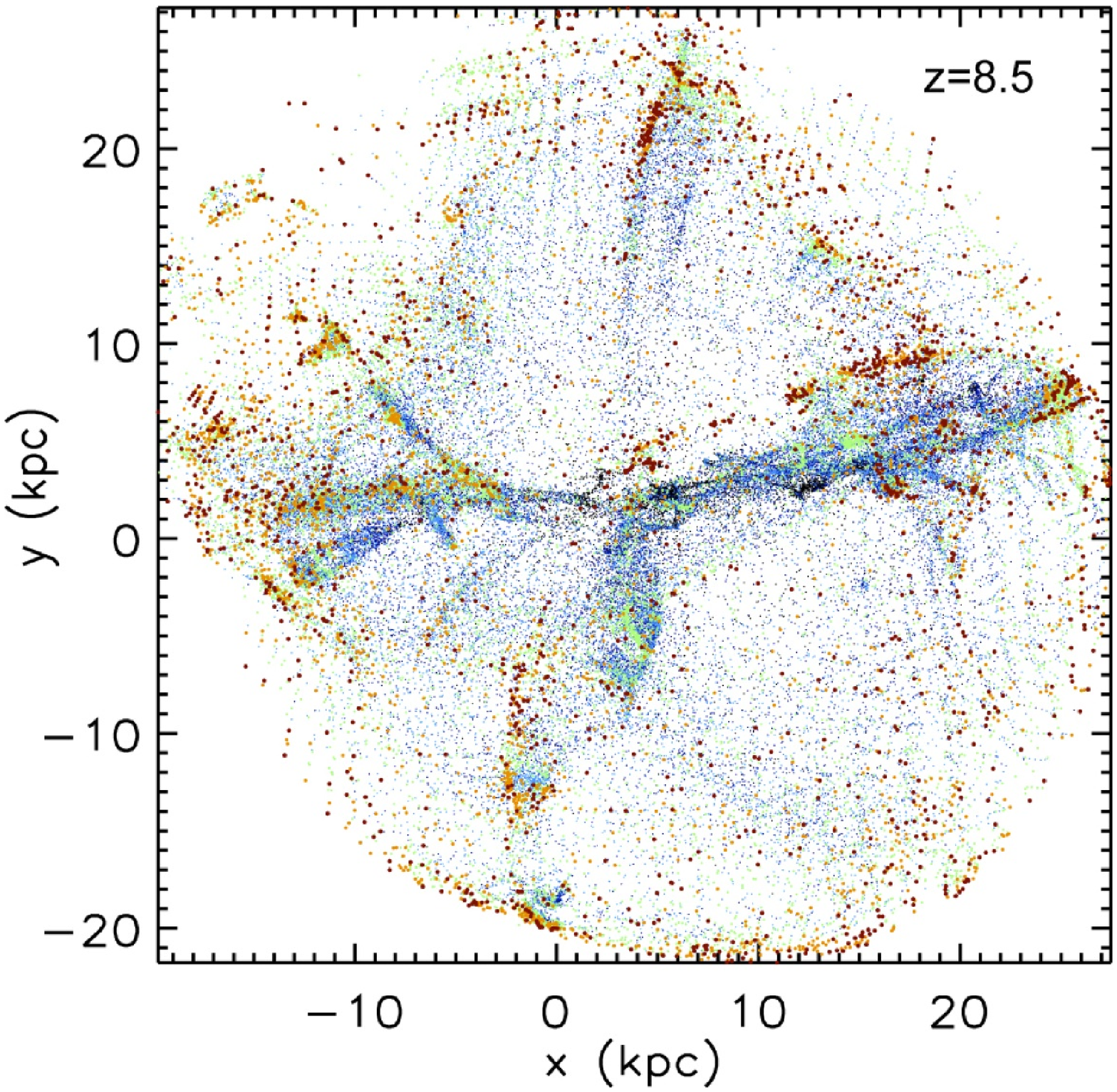}
\includegraphics[width=5.75cm]{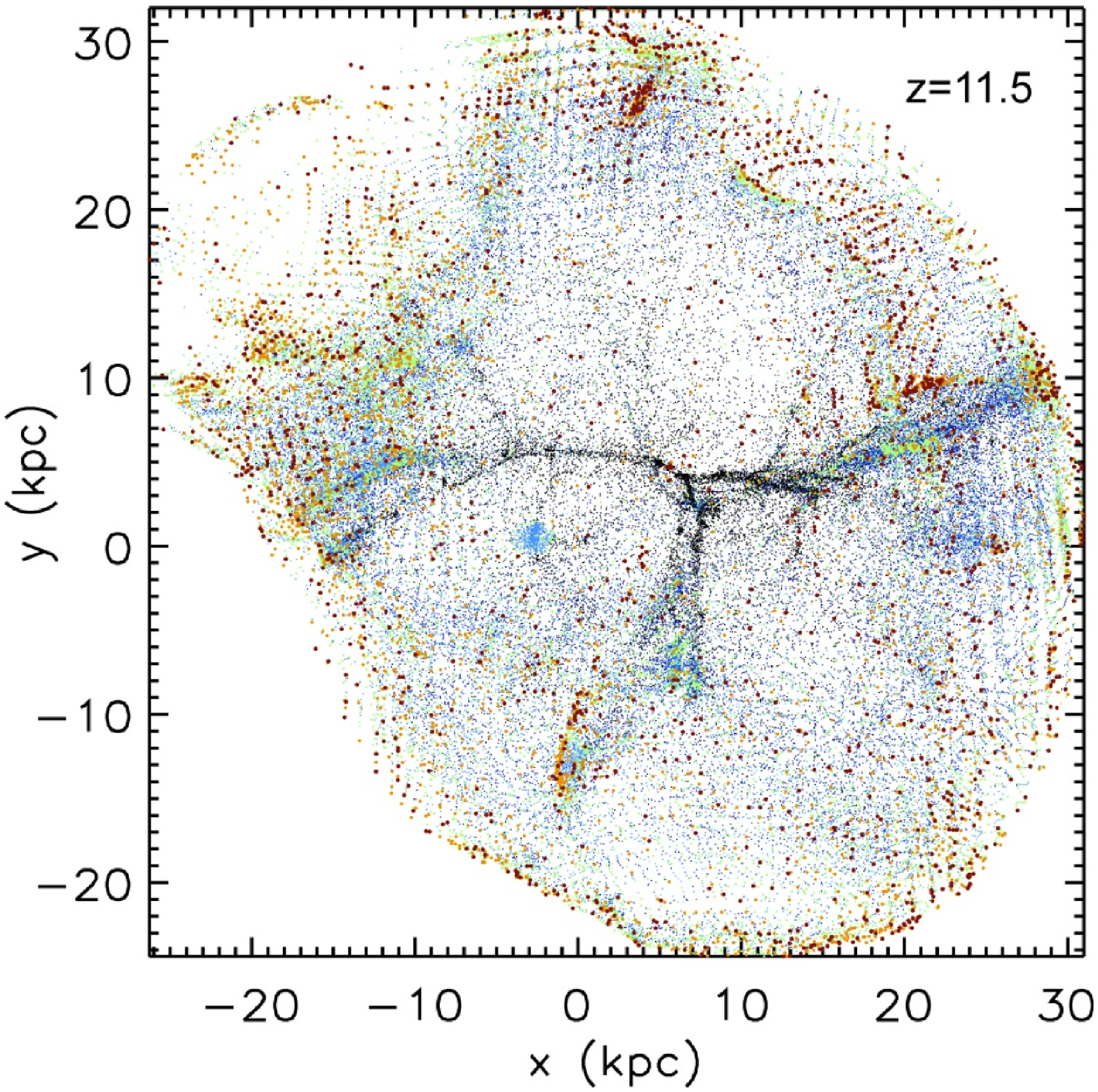}
\includegraphics[width=5.75cm]{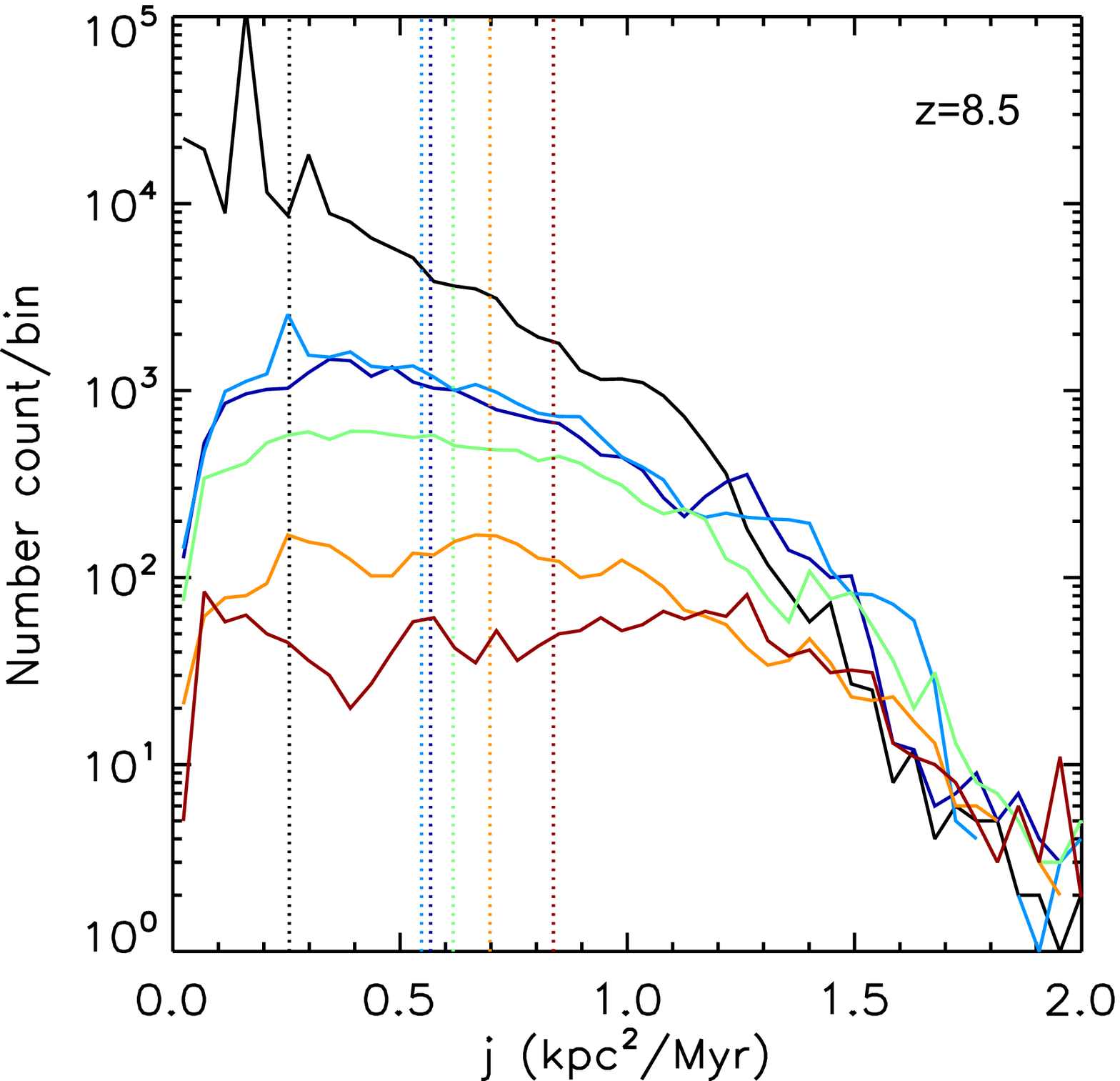}
\includegraphics[width=5.75cm]{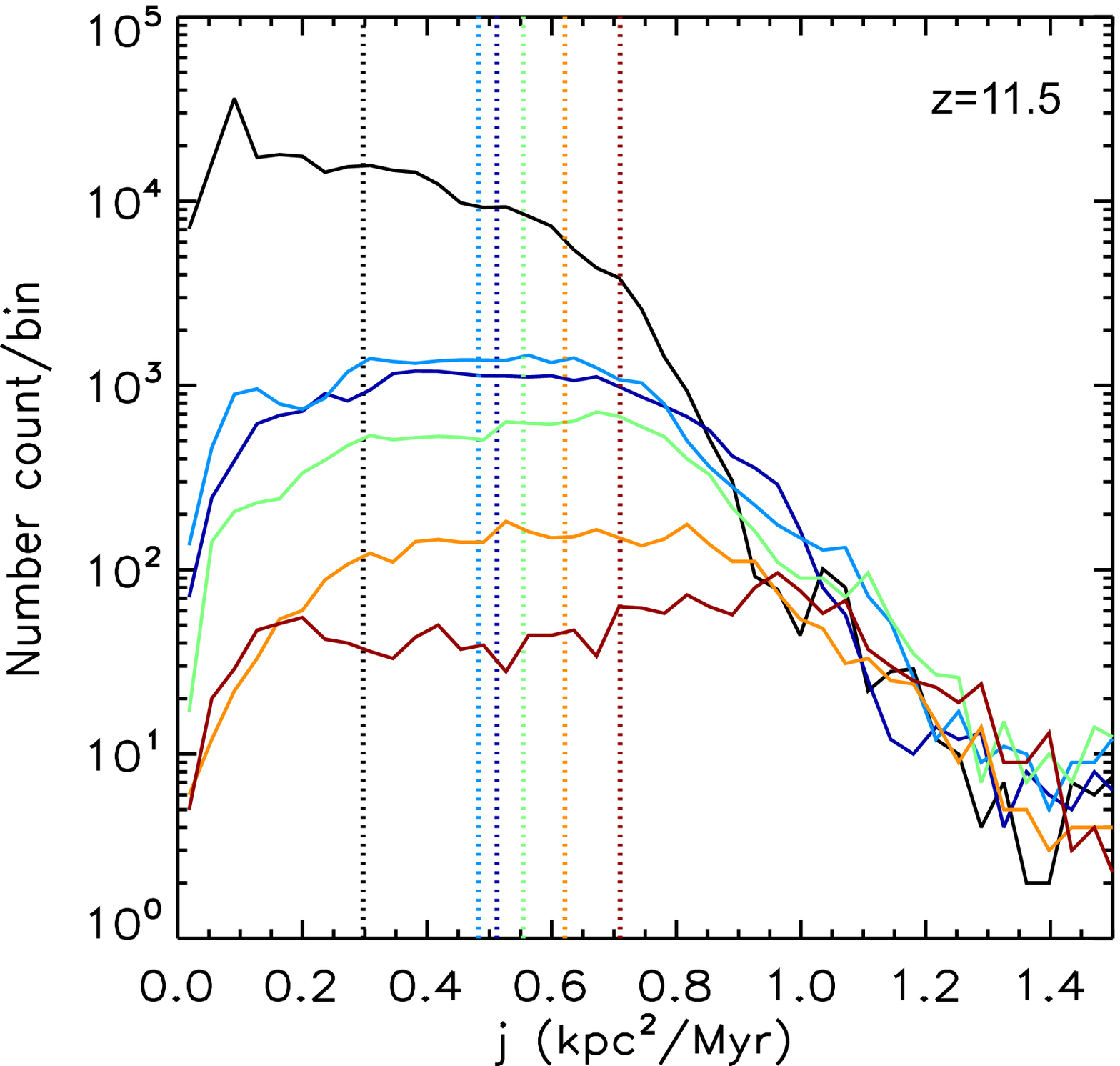}
\includegraphics[width=5.75cm]{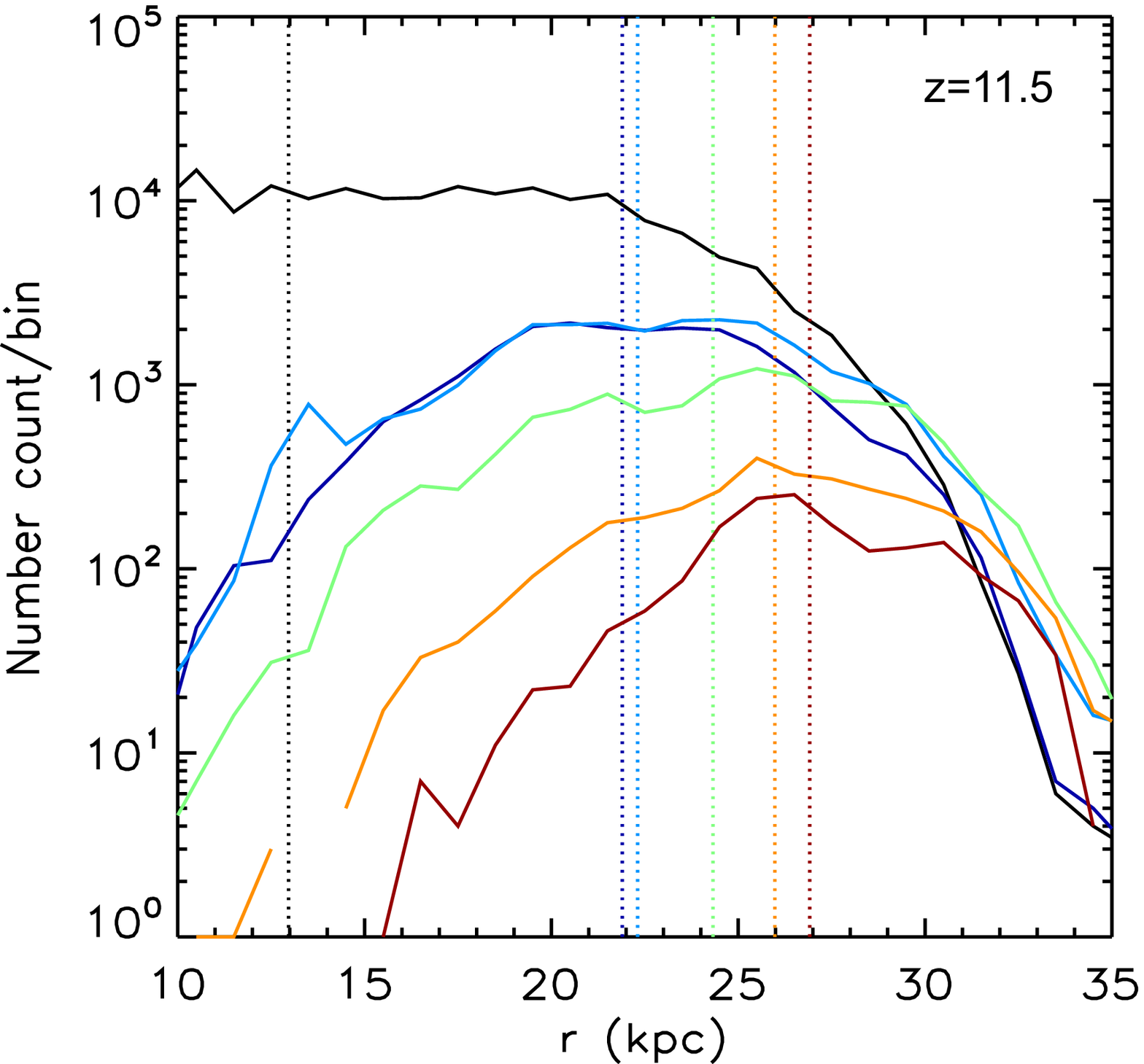}
\caption{  The position of the tracer particles superimposed on the gas temperature at redshift $z=6.5$ 
{\sl (top left panel)}, and the positions of the tracer particles at $z=8.5$ {\sl (top middle panel)} and $z=11.5$ {\sl (top right panel)}. The different colours (black, dark blue, light blue, green, orange, red) for the tracer particles only encode information about their radial position at $z=6.5$ (the concentric, non-overlapping spherical shells they belong to). Note how well the tracer particles follow the shocks and prefer the cold phase.
The  {\sl (bottom left panel)} shows a histogram of the particles' angular momenta at $z=8.5$ {\sl (bottom left panel)}, $z=11.5$ {\sl (bottom middle panel)} and radii  {\sl (bottom right panel)}  (the vertical  dashed line represents the mean of each PDF).
The progenitor of the different annuli around the disc is clearly found in the earlier filaments, while the stratification is consistent with an inside out build up of the circumgalactic medium. As expected, the angular momentum of tracer particles increases with cosmic time.
}\label{fig:trace2}
\end{figure*}

To substantiate the view that the gas angular momentum of galaxies is preferentially advected along filaments
and naturally leads to the inside-out build up of thin discs throughout cosmic ages,
we make use of a high resolution hydrodynamical simulation in the
\nut\ suite \citep{powell10}.  The ultimate goal of the \nut\ suite is to quantify the effect of
various physical mechanisms (cooling, supernovae/AGN feedback, magnetic fields, radiative transfer) on the
properties of well resolved galaxies (several grid cells spanning the scale height of the disc) embedded in their cosmological environment.
To achieve this objective, a cubic volume of the Universe 12.5 comoving Mpc on a side with periodic boundary conditions is tiled
with a uniform 128$^3$ coarse grid. Within this volume, an initial high resolution sphere (1024$^3$ equivalent) of about 1 Mpc in radius is defined which encompasses
the Lagrangian region from which the dark matter particles of a Milky-Way size halo at z=0 originate. As the simulation progresses,
up to 10 additional levels of grid refinement are triggered in a quasi Lagrangian manner so as
to keep the size of the smallest grid cell equal to 12 physical pc at all times.
The initial conditions of the simulation were generated using {\texttt{MPgrafic}} \citep{prunet08}
and adopting a WMAP5 cosmology with $\Omega_m=0.258$, $\Omega_{\Lambda}=0.742$,
$h\equiv  H_0/(100 {\rm km s^{-1} Mpc^{-1}}) = 0.72$ \citep{dunkley09}. This yields a dark matter
particle mass of $\simeq 5.5\times 10^4 \msun$ in the high resolution region.
The simulation includes radiative cooling, star formation and a uniform redshift dependent UV background
instantaneously turned on at $z=8.5$ \citep{haardt96}. The gas density threshold for star formation
is chosen to be equal to the Jeans threshold, i.e. $n_{\rm H,th} = 400 ~{\rm cm^{-3}}$
The minimal mass of star particles is therefore $\simeq 2\times10^4\msun$.

We then track the evolution of the gas flow in our Eulerian grid simulation using tracer
particles. These are mass-less particles whose initial positions and number
exactly match those of the DM particles in the simulation. However, as the simulation proceeds, they are simply advected with the local velocity of the gas calculated by means of a Cloud-In-Cell interpolation.
Figure~\ref{fig:trace2} shows such gas tracing particles in the \nut\ simulation at $z=6.5$
and paints them in a different colour depending on which concentric spherical shell they belong to (top left panel).  These spherical shells are centred on the disk 
galaxy and do not overlap. The background image is a projection of the underlying gas temperature. Note that most of the tracer particles are associated with the 
cold gas component, consistent with the result of \citet{powell10} that the galactic gas growth in mass at high redshift is dominated 
by cold filamentary accretion (See also Appendix~\ref{sec:geom}).
The position of these coloured particles is then displayed at redshifts $z=8.5$ and $z=11.5$ in the top middle and right panels,
illustrating that their radial ordering is preserved over significant periods of time. 
The bottom right, middle and left panels present a histogram of the radii of these particles and their angular momenta at $z=8.5$ and $z=11.5$ respectively.
Notwithstanding that the \nut\ simulation only follows the first Gyr of evolution of a galaxy, it is clear from Figure~\ref{fig:trace2}
that, as we argued in this paper, i) the origin of the cold  gas in the disc and its neighbourhood is filamentary, ii) outer shells of gas come from outer regions which
carry more angular momentum and iii) the angular momentum of accreted gas increases with cosmic time.

\subsection{Extensions and perspectives for low redshift discs?}

In this paper, the emphasis was put on the inflow  from the cosmic
environment onto galactic discs. More specifically, we addressed the implication
of angular momentum transport by cold flows
from large scales down to the virial radii of the dark halos hosting these
central disc galaxies.

In doing so, we deliberately overlooked several important issues. First, our work
has shamelessly  ignored the well documented dichotomy  between galaxies located
in dark matter halos above and below the critical halo mass for shock stability
\citep{keres05, dekel06, ocvirk08, brooks09}.
The mass accretion process in these two galaxy populations is known to be
thermodynamically (cold versus hot mode)
and geometrically (connected to multiple filaments or embedded in them)
different and also to  depend sensitively on redshift, to the point that
even the relevance of cold flows at low redshift (below $z=1.5$) has been
questioned.  
The simulations carried in this paper did not address the angular momentum content of the corresponding possibly quasi-spherical cold gas accretion.
Second, feedback processes internal to the central galaxies (massive stars,
supernovae, AGN) may reduce the relevance of anisotropic infall as they may
partially isotropise outflows, which, in turn, could lead to a disruption of the
inflow. Third, along with the cold gas, the cosmic inflow will advect
galaxy/dark halo satellites. The
relative fraction of  induced minor/major, dry/wet merger will certainly affect
the Hubble type of the central object even though the 'cosmic' consistency of
the accretion
reported in this paper should have a statistical bearing on the preferred  direction of satellite infall at $z=0$ \citep{deason11}. As discussed in Section~\ref{sec:circum},
the dynamics {\sl within} the virial radius of the dark matter host halo is also bound to be significantly more complex:  the shorter dynamical time,
dynamical friction and tidal stripping of satellites, torquing from the central triaxial halo and radiative shocking should introduce a non-negligible amount of angular momentum redistribution for
the gas within the circumgalactic region.

In a recent paper, \cite{tenreiro11}, guided by SPH cosmological simulations,  address the mass assembly of massive {\sl ellipticals}
in the framework of the adhesion model \citep{Gurbatov:1989az,kps1990}.
In accordance with this scenario, they also trace back the origin of star forming gas
to neighbouring cold filaments (cf. Section~\ref{sec:NUTtrace}).
On the other hand, they also find that hot intra-halo gas originates from diffuse regions where strong shells crossing is yet to occur.
As their focus is on the progenitors of massive galaxies, which would correspond
to rounder \citep{pichon99} peak patches associated with rarer events,
they measure very little angular momentum advection in the first stage of the collapse.
As we are concerned with less massive, more common disc-like galaxies,
the associated initial density peak is more ellipsoidal and early angular momentum advection is therefore more important.

The extension of TTT to quantify the amount of angular momentum {\sl expected} to be acquired after
$z_{\rm in}$ constitutes a natural follow-up of the current paper. Such a theory could be constructed from a higher resolution description of the density
field in the patch (i.e. which includes features such as ridges and saddle points that appear when the patch is smoothed on scales smaller than its ellipsoidal
representation via the inertial tensor), using theoretical tools like the skeleton \citep{pogo09}. Such an extension involves predicting  the
relative dynamical importance of the different filaments embedded within the patch, both in terms of advected mass and angular momentum. {
One expects that the six filaments typically connected to a given 3D peak
(Pichon et al. {\sl in prep.}) mostly cancel each other out in terms of
net angular
momentum flux as all the voids surrounding a given peak cannot all induce rotation along the same axis. This extension will need to understand filamentary
bifurcations \citep[e.g.,][]{pogo09} inside the peak patch to see why only two
prevail eventually.}
One would then be able to possibly predict ab initio  
 a fraction of the  cosmic history of galactic spin-up.
  Such a theory for the rig of dark
halos should be within reach,  and will be the topic of
further investigation.

\section*{Acknowledgments}
We thank J. Binney,  S. Codis, A. Dekel, M. Haehnelt, J. Magorrian, S. Peirani, S. Prunet and J. F. Sygnet  for useful comments during the course of this work, and the referee for 
his suggestions.
CP acknowledges support from a Leverhulme visiting professorship at the Physics department of the University of Oxford,  and thanks Merton College, Oxford for a visiting
fellowship, and  Lena for her hospitality. DP thanks the French Canada Research Fund.
TK acknowledges support from a Clarendon DPhil studentship.
JD and AS's research is supported by Adrian Beecroft, the Oxford Martin School and STFC.
YD is supported by an STFC Postdoctoral Fellowship.
We also acknowledge support from the Franco-Korean PHC STAR program  and the France Canada Research Fund. The \mn simulation was run on the MareNostrum machine at the Barcelona
Supercomputing  Centre and we would like to warmly thank the staff for their
support and hospitality. The \nut\
simulation presented here was run on the DiRAC facility jointly funded
by STFC, the Large Facilities Capital Fund of BIS and the University
of Oxford. This research is part of the Horizon-UK project.
Let us thank D.~Munro for freely distributing his {\sc \small  Yorick} programming language and opengl interface (available at {\tt http://yorick.sourceforge.net/})
and T.~Sousbie for the persistent skeleton code {\sc \small  DisPerSe}.

\bibliographystyle{mn2e}
\bibliography{refs}

\appendix

\section{Geometry of Gas Accretion}
\label{sec:geom}
\begin{figure*}
   \centering
   \includegraphics[width=17cm]{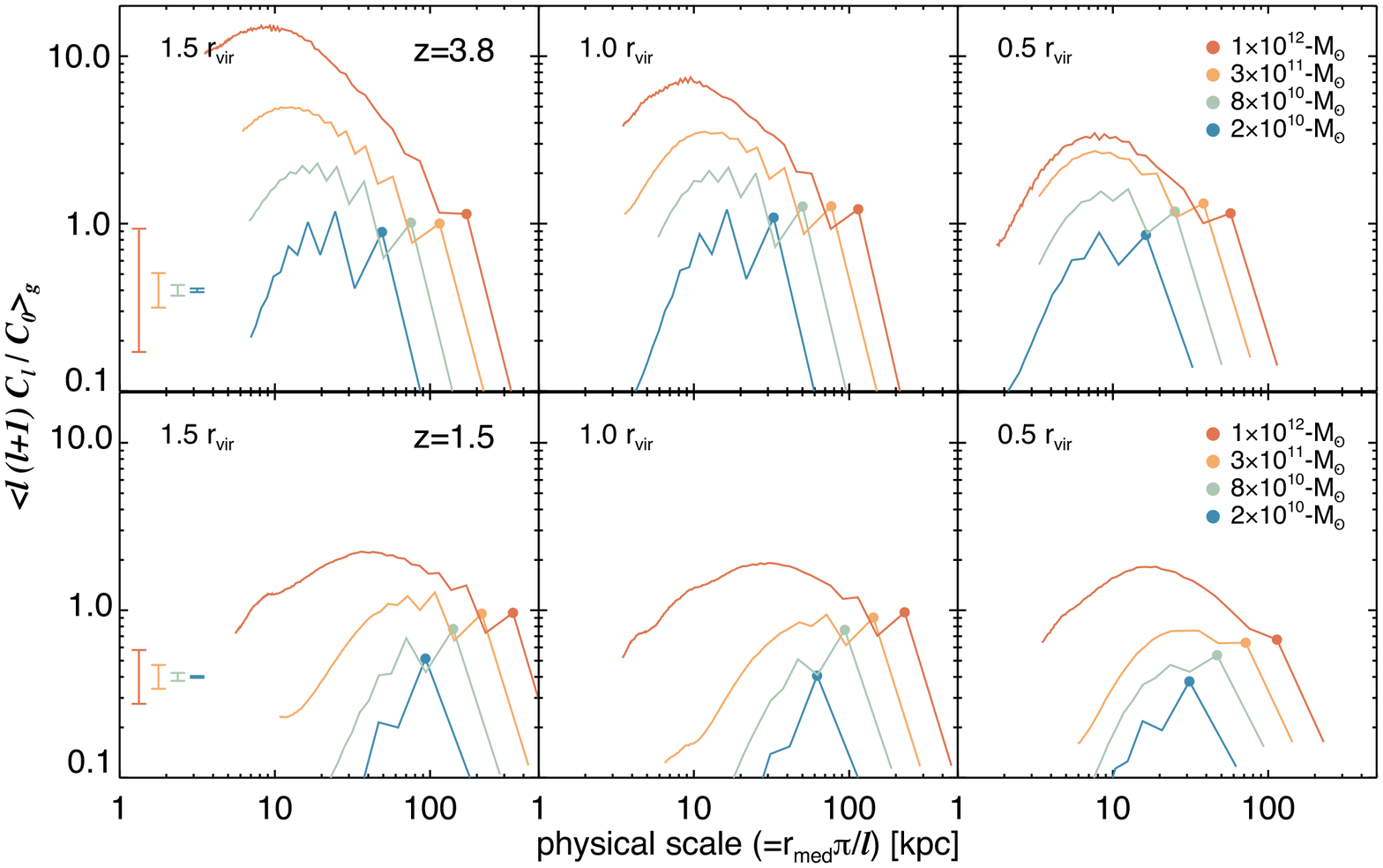} 
   \caption{Normalised power spectrum of gas accretion at different radii. 
  Ordinate indicates the geometric mean of the power spectrum, normalised by the monopole ($C_0$), 
   and abscissa indicates the physical scale ($\equiv r_{\rm med} \pi/l$) corresponding to each multipole ($l$). 
   The median radius ($r_{\rm med}$) of each subsample is used to compute the physical scale. 
   Different colour-coding denotes different halo mass bins, and the quadrupole ($l=2$) is marked as a filled circle.
   Power spectra for high ($z=3.8$) and  low redshift ($z=1.5$) are shown in upper and lower panels, respectively. 
   The radius at which the power spectrum is measured decreases from left to right. 
  Error bars in the left panels are the maximum standard errors of the mean in each mass bin.
   The most massive haloes ($\mvir > 10^{13} \msun$) are not included in this analysis 
   because of their small number. This figure shows that  statistically, the advected gas is coming along 
   filaments whose cross section increases with cosmic time and for lighter halos.
   }
   \label{fig:power_prad}
\end{figure*}

The scenario proposed in Section 3 is based on the assumption that cold filamentary accretion 
primarily accounts for the growth of the central gaseous disc at high redshift. 
This has been studied by several authors \citep[e.g.,][]{brooks09, kimm11}. In this section, 
we aim to provide further statistical evidence  for the geometrical nature of the infall
by looking at the power spectrum of gas accretion from the \mn\ simulation.

We make use of {\sc HEALPiX} package \citep{hivon05} to transform accretion maps 
into power spectra, $l(l+1)C_l$, with the tessellation parameter {\sc NSIDE}$=128$. 
We first divide the sample by halo mass and compute the average power spectra for each mass bin.
We average the power spectra normalised by the monopole term, i.e. $\left<l(l+1)C_l/C_0\right>_g$, 
where $\left<\,\,\,\right>_g$ denotes the geometrical mean (in order to properly account for the typically log-normal distribution of the different power spectra of individual halos). 
Note that $C_0$ represents the global accretion rate, and hence the power normalised 
by $C_0$ corresponds to the relative contribution from each mode $l$. 
In order to account for the limited resolution of the simulation, 
the measurement of the power spectra for modes greater than $l_{\rm lim}\equiv \pi\Delta x_{\rm min}/r$ are discarded, 
where $\Delta x_{\rm min}$ is the minimum size of the AMR cell used to obtain the map, and $r$ is the 
radius of  the corresponding spherical shell. 

Figure \ref{fig:power_prad} shows the power spectrum of the gas accretion as a function of {\sl  
physical scales} corresponding to each harmonic multipoles at $z=3.8$ (upper panels). 
The virial radii vary with  halo mass, so each multipole does not correspond to the 
same physical size for all dark matter haloes. Thus, we use  the median radius of each subsample
$r_{\rm med}\pi/l$, to investigate the physical scale of the accretion.
We plot the results when more than 95 per cent of the halos show a 
reliable estimate of the power spectrum. This induces a cutoff of $\simeq$2--10 kpc 
depending on halo mass.

We find that the characteristic scale of gas accretion (corresponding to the peak of the power spectra) is around 10--30 kpc at $z=3.8$,
which is roughly the scale of filaments in Figure \ref{fig:visual}. 
This supports the view that gas accretion is mainly filamentary at high redshift.
At lower redshift ($z=1.5$) one expects that the filaments become more diffuse and broader
as the universe expands. This view is borne out by the bottom panels of Figure \ref{fig:power_prad}
which suggest that the characteristic size of the accretion is 
$\sim 60$ kpc at the virial radius for z=1.5. In particular, for less massive halos ($M_{\rm vir}\leq 5\times 10^{11}\msun$) 
that are likely to end up forming spiral galaxies, the power spectrum is dominated by the dipole term, 
implying that two filaments are responsible for the accretion.
It is worth noting that at larger distances ($r=1.5\, r_{\rm vir}$) 
the characteristic scale becomes larger than the diameter of $5\times10^{10}\msun$ halos ($\sim 80$ kpc), 
indicating that smaller halos are probably embedded within the filament(s). 

An interesting feature of Figure \ref{fig:power_prad} is the oscillation  in the power spectra. 
Such an oscillating power spectrum at {\em low} multipoles can be found when the contribution from 
even modes is superior to that from odd modes.  
The preference for even modes indicates that gas accretion is dominated by an even
 number of sources, which are likely to be separated by $\sim \pi$ on the sky. 
 If the accretion were dominated by one strong source, 
 i.e. a satellite galaxy, the power spectrum for odd modes would be larger than that for even modes 
 and the oscillatory feature would vanish or be reversed.  
 However, it should be noted that the larger contribution from even modes does not 
 necessarily mean that any even number of infalling structures will produce the same signal. 
 Only when the contribution from such structures is significantly larger than that of other sources and
 when these structures are separated by $\sim \pi$ does the power spectrum show this oscillatory feature.
Note finally that the oscillatory feature turns out to be more significant for less massive haloes,
 suggesting again that the accretion is highly likely to be bipolar for these haloes.

\section{Angular momentum orientation correlation along filaments}
\label{sec:space}

\begin{figure}
 \includegraphics[width=8.5cm]{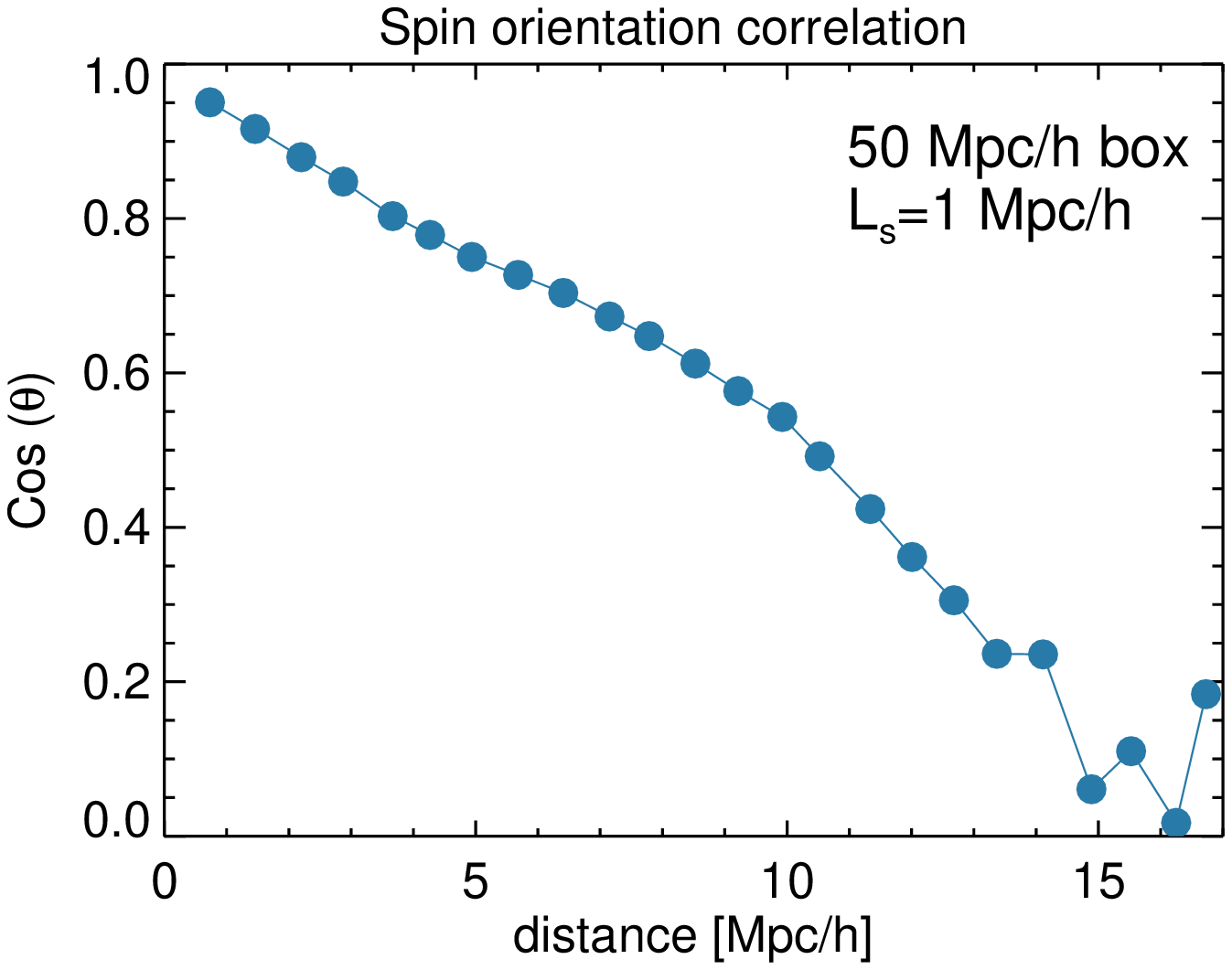}
\caption{Expectation of the relative orientation of the angular momentum of the filamentary flow as a function of the separation along
the filament; this quantity is found to be quite stable w.r.t. redshift.
}\label{fig:space-correl}
\end{figure}

In Section~\ref{sec:hydro} we found that the orientation of advected angular momentum was fairly stationary w.r.t. cosmic time. Let us
underpin this measurement by investigating the coherency of this orientation along the large scale structure filaments.
Consider a set of filaments (defined here as a set of segments of the skeleton between two peaks, see \cite{sousbie09}).
Let $s_i$ and $s_j$ be the curvilinear coordinates along that filament  of two segments and $\Delta \theta_{ij}$ be the relative angle
of the two angular momenta , $\mathbf{J}_i$ and $\mathbf{J}_j$ separated by $s_i-s_j$ (i.e. $ \cos (\Delta \theta_{ij}) = \mathbf{J}_i \cdot \mathbf{J}_i/|J_i||J_j|$).
Let us define the expectation, $\langle \cos(\theta)\rangle(r) $, of the relative angle between the angular momentum orientation as a function of distance along the filament
as
\[
\langle \cos(\theta)\rangle (r)= \sum_{i,j | \,\,\, | s_i-s_j|=r\pm \Delta r}  \cos(\Delta \theta_{ij}) \Big/  \sum_{i,j | \,\,\, | s_i-s_j|=r\pm \Delta r} 1\,,
\]
where the summation is over all pairs, $i,j$ belonging to the same filament for which the separation in curvilinear coordinate falls within $\Delta r$ of $r$.
The average of this expectation over all filaments in the simulations is shown on Figure~\ref{fig:space-correl} for the skeleton of a dark matter simulation of $256^3$ particles in a cube of  volume (50 Mpc$/h)^3$ smoothed over a
1Mpc$/h$ scale. The orientation of the spin of dark matter particles is clearly correlated over scales of the order of 10 Mpc$/h$.
This result is qualitatively consistent with the temporal correlation of Figure~\ref{fig:cosinePDF}.

\vfill

\end{document}